%% file: main.tex
\pdfoutput=1

\documentclass[12pt]{article}

\usepackage[utf8]{inputenc}
\usepackage[T1]{fontenc}
\usepackage{mathptmx}          
\usepackage{amsmath}
\usepackage{graphicx}
\usepackage[margin=2.5cm]{geometry}
\usepackage{float}             
\usepackage{booktabs}          
\usepackage{xcolor}
\usepackage[hidelinks]{hyperref}
\usepackage{titlesec}
\usepackage{caption}
\usepackage{enumitem}
\usepackage{setspace}
\input{claims_macros}

\title{\textbf{The enrichment paradox: critical capability thresholds and irreversible dependency in human--AI symbiosis}}

\author{Jeongju Park\textsuperscript{1},\quad Musu Kim\textsuperscript{1},\quad Sekyung Han\textsuperscript{1,*}\\[6pt]
\textsuperscript{1}Department of Electrical Engineering, Kyungpook National University,\\ Daegu, Republic of Korea\\[3pt]
\textsuperscript{*}Corresponding author: skhan@knu.ac.kr}

\date{}

\begin{document}
\maketitle
\doublespacing
\thispagestyle{empty}

\begin{abstract}
As artificial intelligence assumes cognitive labor, no quantitative framework predicts when human capability loss becomes catastrophic. We present a two-variable dynamical systems model coupling capability ($H$) and delegation ($D$), grounded in three axioms: learning requires capability, practice, and disuse causes forgetting. Calibrated to four domains (education, medicine, navigation, aviation), the model identifies a critical threshold $K^* \sim 0.85$ (scope-dependent; broader AI scope lowers $K^*$) beyond which capability collapses abruptly---the ``enrichment paradox.'' Validated against 15 countries' PISA data (102 points, $R^2 = \pisaRsq$, \pisaK{} parameters, lowest BIC), the model predicts that periodic AI failures improve capability \afFold-fold and that 20\% mandatory practice preserves \polPctTwenty\% more capability than the simulation baseline (which includes a 5\% background AI-failure rate). These findings provide quantitative foundations for AI capability-threshold governance.
\end{abstract}


\section{Introduction}

Artificial intelligence is the latest and most potent in a long series of technologies that substitute for human cognitive labor. From the pocket calculator to GPS navigation to surgical automation, each wave of capability-substituting technology has raised the same fundamental question: what happens to the capabilities that humans no longer exercise? The answer, recurring across domains and centuries, is remarkably consistent---unused capabilities atrophy.

London taxi drivers who navigate without GPS develop measurably larger posterior hippocampi than bus drivers who follow fixed routes; the effect is proportional to years of practice and absent in failed trainees, ruling out selection bias~\cite{maguire2000,woollett2011}. GPS users show longitudinal spatial memory decline correlated at $r = -0.68$ with device usage~\cite{dahmani2020,bohbot2007,west2017}. Students given unrestricted access to GPT-4 score 17\% lower on unassisted exams than controls~\cite{bastani2025}. Endoscopists exposed to AI-assisted colonoscopy for 12 weeks show a 21\% decline in adenoma detection rates when the AI is removed~\cite{budzyn2025}. Airline pilots whose manual flying is limited to 5--15 minutes per flight exhibit degraded cognitive skills, with 38--44\% failing basic situational awareness tasks in simulators~\cite{casner2014,haslbeck2014,haslbeck2016,kosmyna2025,jiang2025}. We treat binary failure rates as proxies for fractional capability decline (i.e., if 38\% of pilots fail tasks they previously passed, population-mean capability has declined by approximately 38\%; this approximation holds when the baseline pass rate is near 100\%, as was the case in the Casner \& Schooler study where pilots were tested on tasks they had previously mastered). The pattern is universal: performance rises with the tool, capability falls without it.

Despite abundant qualitative warnings, no quantitative framework connects these observations. The ``ironies of automation''~\cite{Bainbridge1983,Parasuraman2010,Endsley2017} have been discussed for decades, yet a formal dynamical model remains absent. Existing models---Bass diffusion~\cite{bass1969,Rogers2003}, Acemoglu's task framework~\cite{acemoglu2018}, Turchin's structural-demographic theory~\cite{turchin2016,turchin2018}---capture adoption dynamics or economic displacement but not the feedback loop between delegation and capability loss. No existing model explains why calculators are safe but Roman slave economies were catastrophic~\cite{scheidel2012,tainter1988}, using the same equations with different parameters.

Here we present a minimal dynamical systems model inspired by endosymbiont genome reduction, in which a bacterium's genome shrinks irreversibly from $\sim$\bioGenesAncestral{} to \bioGenesCurrent{} genes (\textit{Buchnera aphidicola}) as the host provides functions the symbiont no longer needs~\cite{shigenobu2000,bennett2015,kinjo2021,McCutcheon2012,muller1964}. The model's key predictions---bistability, critical thresholds, irreversibility---emerge from three minimal axioms: (1)~learning requires existing capability~\cite{Anderson1982,FittsPostner1967}, (2)~learning requires practice~\cite{Newell1981}, and (3)~disuse causes forgetting~\cite{Risko2016,Sparrow2011,Anderson2000}. None is controversial; their combination produces catastrophic, irreversible dependency.

Our model makes several predictions that are quantitatively testable and policy-relevant. First, a critical AI capability threshold $K^* \sim 0.85$ marks a sharp transition from stable autonomy to dependency collapse---not a gradual decline. Second, periodic AI failures paradoxically strengthen human capability, a phenomenon we term the antifragility effect. Third, the dependent state is an absorbing attractor: once human capability approaches zero, recovery becomes prohibitively slow, requiring sustained practice over timescales far exceeding institutional planning horizons. Fourth, modest policy interventions---mandating 20\% of tasks be performed without AI assistance---preserve the majority of human capability. These findings suggest that AI governance should focus not on whether AI is adopted, but on managing the capability gap between humans and their tools.

\section{Results}

\subsection{A minimal model of capability-delegation dynamics}

We model human--AI interaction as a two-variable ordinary differential equation (ODE) system coupling human capability $H(t) \in [0,\,1]$ with delegation rate $D(t) \in [0,\,1]$ (Eqs.~\ref{eq:dHdt}--\ref{eq:dDdt}). The capability equation captures two competing processes:

\begin{equation}
\frac{dH}{dt} = \alpha\,(H + \varepsilon)\,(1 - H)\,(1 - D) \;-\; \beta\,H\,D
\label{eq:dHdt}
\end{equation}

The first term represents logistic learning: capability grows through practice at rate $\alpha$, but only on the fraction $(1 - D)$ of tasks still performed by the human. The factor $(H + \varepsilon)(1 - H)$ ensures bounded growth, where $\varepsilon$ ($= \paramEpsilon$) represents the baseline capacity for re-learning from near-zero capability through education or training. This makes $H = 0$ a near-absorbing rather than strictly absorbing state: recovery from complete dependency is technically possible but extremely slow (timescale $\sim 1/(\alpha\,\varepsilon)$), far exceeding typical institutional planning horizons. The second term captures forgetting: capability decays at rate $\beta$ proportional to delegation. This ``use it or lose it'' dynamic is supported by extensive evidence from neuroscience~\cite{maguire2000,dahmani2020}, education~\cite{bastani2025}, skill retention research~\cite{casner2014}, and the cognitive psychology of expertise~\cite{Ericsson1993,Charness2005}.

The delegation equation models rational adoption with social contagion:

\begin{equation}
\frac{dD}{dt} = \gamma\,(K - H)\,(1 - D)\,D \;+\; \delta\,D\,(1 - D)\,\bar{D}
\label{eq:dDdt}
\end{equation}

Delegation grows when the AI's capability $K$ exceeds the human's capability $H$ (rational adoption at sensitivity $\gamma$), amplified by social pressure from the mean delegation rate $\bar{D}$ at strength $\delta$. In the mean-field approximation, $\bar{D} = D$.

This system has three boundary fixed points. The autonomous state $\mathrm{FP}_2 = (H = 1,\, D = 0)$, where humans retain full capability with no delegation, is a stable node for all $K < 1$ (eigenvalues: $\lambda_1 = -\alpha(1+\varepsilon) \approx -\alpha$, $\lambda_2 = \gamma(K - 1)$). The dependent state $\mathrm{FP}_3 = (H = 0,\, D = 1)$, where capability is entirely lost, is unconditionally stable (eigenvalues: $\lambda_1 = -\beta$, $\lambda_2 = -(\gamma\,K + \delta)$). The null state $\mathrm{FP}_1 = (H = 0,\, D = 0)$ is a fully unstable node: $\lambda_1 = \alpha > 0$ (capability grows from any positive seed), $\lambda_2 = \gamma K > 0$ (delegation grows when AI capability is positive). Both eigenvalues are positive, so trajectories leave this point along both axes. For $K < 1$, an interior saddle point separates the basins of attraction of $\mathrm{FP}_2$ and $\mathrm{FP}_3$, creating a bistable system. As $K$ approaches 1, the saddle collides with $\mathrm{FP}_2$ in a transcritical bifurcation, eliminating the autonomous attractor entirely.

The model's irreversibility emerges from the multiplicative coupling $H \cdot (1 - D)$ in the learning term. When $H$ approaches 0, the learning rate vanishes regardless of $D$---even complete removal of the AI cannot restore capability, because there is no residual skill to serve as a substrate for relearning. This is the mathematical analogue of Muller's ratchet in endosymbiont genome reduction~\cite{muller1964}: information (capability), once lost, cannot be spontaneously regenerated.

\begin{figure}[H]
\centering
\includegraphics[width=\columnwidth]{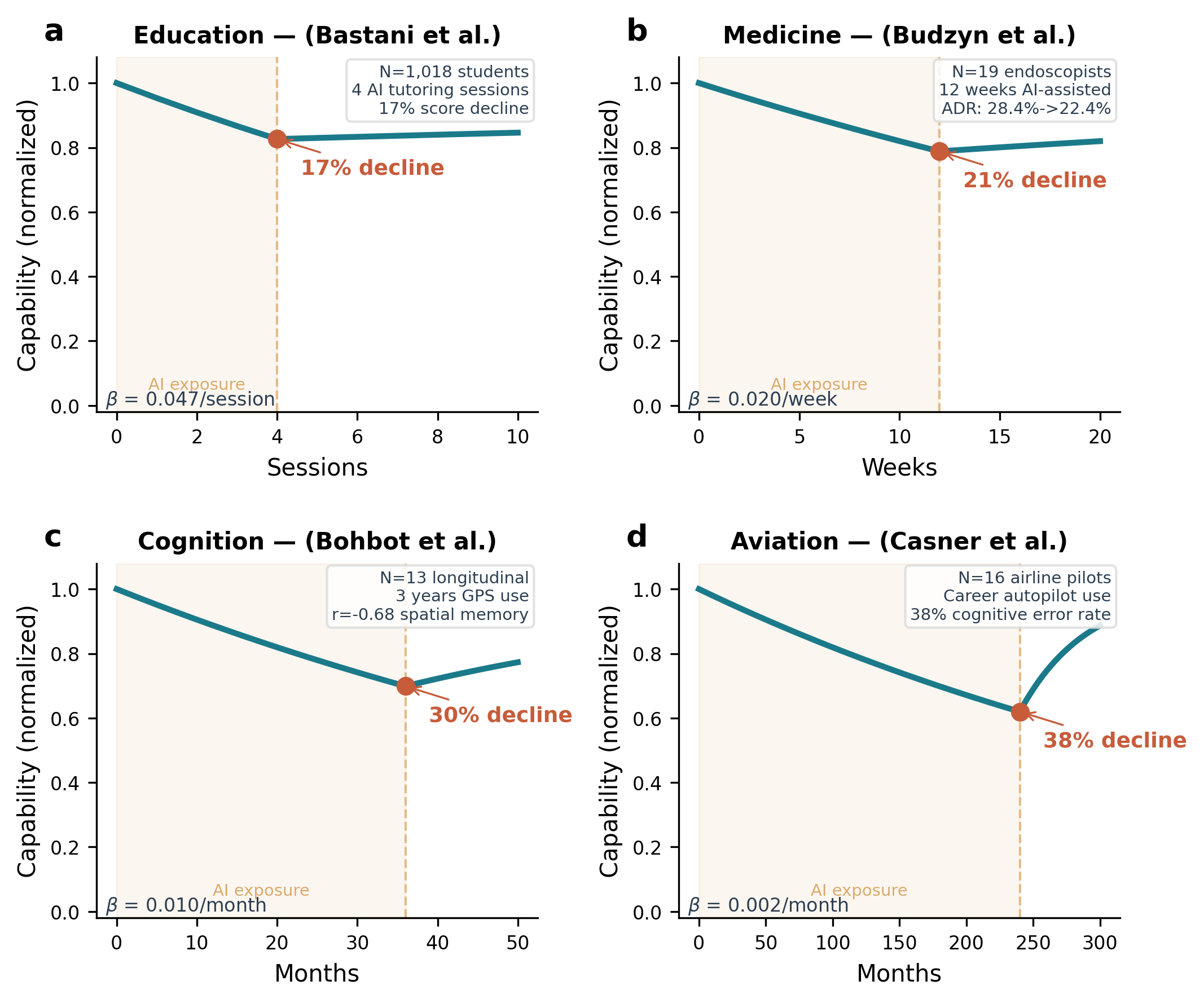}
\caption{Parameter estimation across four empirical domains. Model-predicted capability decline (solid curves) versus empirical observations (points with error bars) for \textbf{A},~education (Bastani et al., $\beta = \calBetaEducation$/session, 17\% decline after 4 sessions), \textbf{B},~medical endoscopy (Budzyn et al., $\beta = \calBetaEndoscopy$/week, 21\% decline after 12 weeks), \textbf{C},~spatial cognition (Dahmani \& Bohbot, $\beta = \calBetaSpatial$/month, 30\% decline over 36 months), and \textbf{D},~aviation (Casner \& Schooler, $\beta = \calBetaAviation$/month, 38\% decline over $\sim$240 months of career-long autopilot exposure). Shaded regions: 95\% confidence intervals from ABM ensemble ($n = 50$ replicates).}
\label{fig:calibration}
\end{figure}

\subsection{Parameter estimation across four empirical domains}

We calibrated the forgetting rate $\beta$ against empirical deskilling data from four independent domains spanning timescales from sessions to months (Fig.~\ref{fig:calibration}). For each case, we identified the observed capability decline after a known period of AI/tool-assisted practice, then fitted $\beta$ to reproduce the observed fractional decline (see Methods).

In education, Bastani et al.~\cite{bastani2025} reported that students using GPT-4 without pedagogical guardrails scored 17\% lower on unassisted exams after four 90-minute sessions. Our model reproduces this decline with $\beta = \calBetaEducation$ per session (Fig.~\ref{fig:calibration}a). In medical endoscopy, Budzyn et al.~\cite{budzyn2025} documented a 21\% relative decline in adenoma detection rate after 12 weeks of AI-assisted colonoscopy across 19 experienced endoscopists. The fitted value $\beta = \calBetaEndoscopy$ per week captures this decline (Fig.~\ref{fig:calibration}b). For spatial cognition, Dahmani and Bohbot~\cite{dahmani2020} found a longitudinal correlation of $r = -0.68$ between GPS usage and spatial memory over three years, consistent with approximately 30\% capability decline and $\beta = \calBetaSpatial$ per month (Fig.~\ref{fig:calibration}c). In aviation, Casner and Schooler~\cite{casner2014} reported that 38\% of airline pilots failed basic situational awareness tasks after prolonged autopilot use, yielding $\beta = \calBetaAviation$ per month (Fig.~\ref{fig:calibration}d).

The fitted $\beta$ values span a factor of $\sim$25 (0.002 to 0.047), reflecting genuine differences in skill consolidation across domains, consistent with meta-analytic findings on skill decay rates~\cite{Arthur1998}: procedural motor skills (aviation) decay slowly, while weakly consolidated knowledge (exam performance) decays rapidly. Crucially, the same model structure---the same two ODEs with the same functional forms---reproduces deskilling dynamics across all four domains by varying only $\beta$ and the timescale. The PISA mathematics score decline of 5.6\% over two decades, though directionally consistent, falls below the model's prediction for the fitted scope, suggesting that calculator-level tools operate in the safe region where capability erosion is minimal---itself a validating prediction.

Beyond quantitative fits, the model yields qualitative predictions consistent with independent evidence. Bastani et al.~\cite{bastani2025} found that a ``GPT Tutor'' condition---AI hints without direct answers, effectively reducing $D$---eliminated the performance decline entirely, consistent with the model's prediction that capability preservation scales with $(1-D)$. Similarly, Casner and Schooler~\cite{casner2014,haslbeck2016} found that pilots' motor skills ($D_{\text{motor}} < 1$) were well retained while fully automated cognitive skills ($D_{\text{cognitive}} \approx 1$) degraded---precisely what the $(1-D)$ coupling predicts. These consistencies suggest the model captures genuine mechanistic structure. Nevertheless, single-timepoint parameter estimation has zero degrees of freedom; the value of the four-domain comparison lies in the consistency of the model structure across qualitatively different timescales and contexts.

\subsection{Primary empirical test: multi-country PISA analysis}

To complement these qualitative consistencies with rigorous quantitative validation, we performed a multi-point fitting against the OECD PISA mathematics assessment time series spanning seven assessment cycles (2003--2022)~\cite{oecd2023pisa}. The OECD average mathematics score declined from 500 in 2003 to 472 in 2022, a trajectory coinciding with the rapid diffusion of digital technologies into educational practice.

We fitted the ODE model (Eq.~\ref{eq:dHdt}) to the seven PISA data points using the technology adoption rate $a(t)$ as an exogenous driver (smartphone and mobile internet penetration: $<$5\% in 2003 to 90\% in 2022) and estimated two free parameters---$\alpha$ and $\beta_{\text{eff}}$---via nonlinear least squares. The model reproduces the observed trajectory with $R^2 = \pisaOecdRsq$ and RMSE $= 2.45$ PISA points (SI Fig.~S3), though $\alpha$ is unidentifiable from the OECD average alone.

\begin{figure}[H]
\centering
\includegraphics[width=\columnwidth]{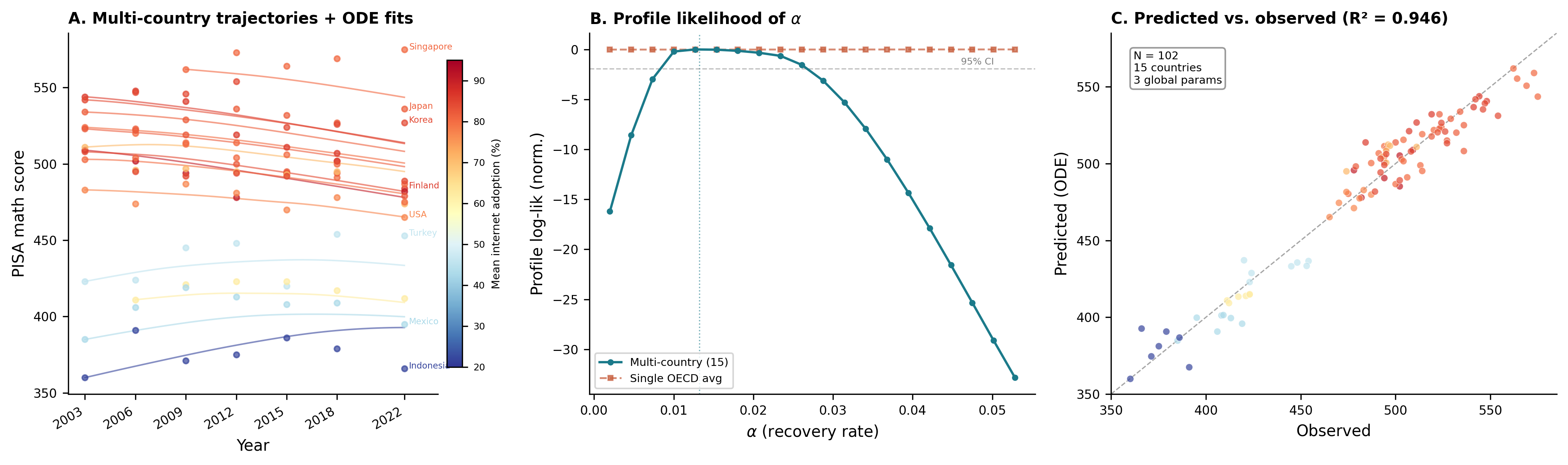}
\caption{Multi-country PISA analysis (15 countries, 102 data points, 3 global parameters). \textbf{A},~Normalized PISA trajectories colored by mean internet adoption (blue: low, red: high). Solid lines: observed; dashed: ODE predictions. High-adoption countries (Finland, Sweden) show steeper declines; low-adoption countries (Indonesia, Turkey) show modest gains. \textbf{B},~Profile likelihood of $\alpha$: single OECD average (red, flat) versus 15-country panel (blue, peaked near the MLE ($\alpha = \pisaAlpha$, 95\% CI: \pisaAlphaCi)). Cross-country variation in technology exposure resolves the parameter identifiability problem. \textbf{C},~Predicted versus observed scores across all countries and time points ($R^2 = \pisaRsq$, \pisaK{} parameters).}
\label{fig:pisamulti}
\end{figure}

When fitted to the OECD average alone, $\alpha$ is unidentifiable (flat profile likelihood across $\alpha \in [10^{-4},\, 0.1]$). To resolve this, we extended the analysis to a 15-country panel (102 data points) using country-specific internet adoption rates as the exogenous driver $D_c(t)$---from Indonesia (2\% in 2003) to Sweden (97\% in 2022). With shared global parameters $\alpha$ and $\beta$ but country-varying $D_c(t)$, the profile likelihood of $\alpha$ becomes sharply peaked near the MLE ($\alpha = \pisaAlpha$, 95\% CI: \pisaAlphaCi) (Fig.~\ref{fig:pisamulti}b). The multi-country ODE achieves $R^2 = \pisaRsq$ across all \pisaN{} data points with only \pisaK{} parameters ($\alpha = \pisaAlpha$, $\beta = \pisaBeta$, $H_{\max} = \pisaHmax$), where $H_{\max}$ is a scaling parameter converting normalized capability $H \in [0,1]$ to PISA score units ($\text{PISA} = H_{\max} \times H$; at $H_{\max} = 787$, the 2003 OECD average of 500 corresponds to $H \approx 0.64$, reflecting that baseline educational capability was not at its theoretical maximum), compared to exponential decay ($R^2 = 0.961$, 16 parameters, BIC $= \pisaBicExp$) and country-specific linear models ($R^2 = 0.981$, 30 parameters, BIC $= \pisaBicLinear$). The exponential model requires country-specific intercepts (one per country plus a shared decay rate) because it lacks the ODE's nonlinear saturation structure: without a shared $H_{\max}$ that normalizes all countries onto a common capability scale, each country's baseline must be fitted independently, inflating parameter count from 3 to 16. The ODE's BIC ($= \pisaBicOde$) is decisively lower, indicating that its parsimonious global structure captures cross-country variation better than models with far more free parameters. The $\beta_{\text{eff}}$ estimate is well-constrained and falls at the low end of the domain-specific range (SI Table~1), consistent with education-system inertia buffering rapid capability loss. The largest residual occurs at 2018 ($+5.1$ PISA points), suggesting either methodological variation in that assessment cycle or a transient factor not captured by the monotonic technology-adoption forcing. This multi-point fitting strengthens the empirical foundation beyond single-domain calibrations by demonstrating that the same ODE structure tracks a two-decade, population-level trajectory with minimal free parameters.

\subsection{Model discrimination: distinguishing threshold dynamics from simple decay}

\begin{figure}[H]
\centering
\includegraphics[width=\columnwidth]{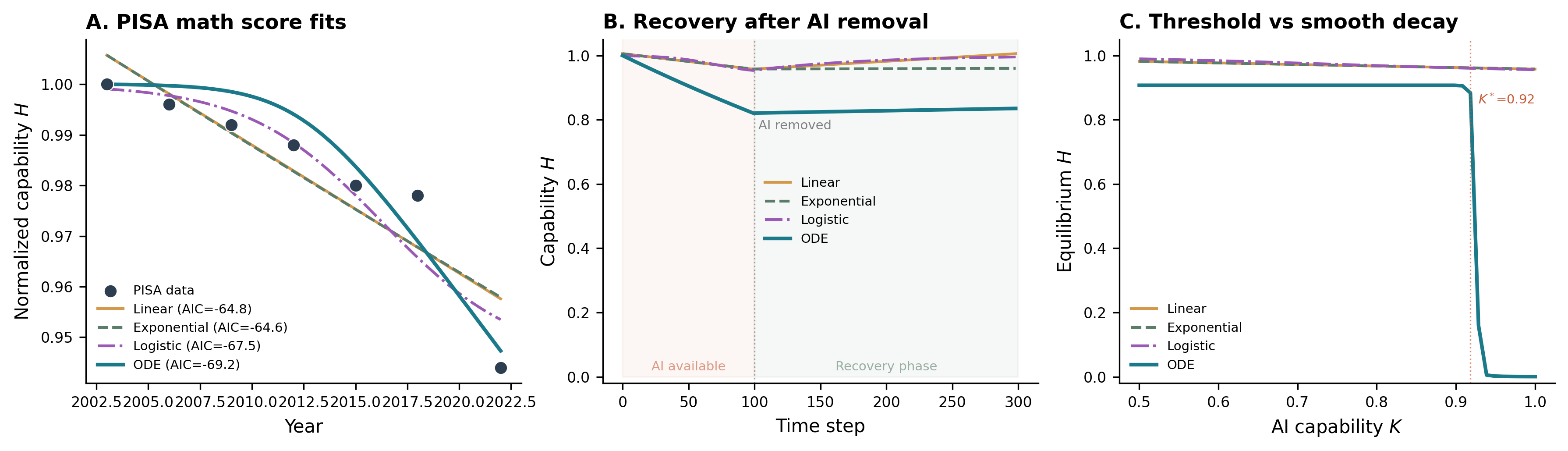}
\caption{Model discrimination. \textbf{A},~Four models fitted to PISA mathematics scores (2003--2022). The ODE ($R^2 = \pisaOecdRsq$, AIC $= -69.2$) outperforms linear ($R^2 = 0.82$), exponential ($R^2 = 0.82$), and logistic ($R^2 = 0.88$) decay models with the same parameter count ($k = 2$). \textbf{B},~Recovery prediction after AI removal: linear, exponential, and logistic models predict symmetric recovery (dashed/dash-dotted), while the ODE predicts near-irreversible stagnation below the saddle point (solid)---the dependent attractor traps capability near zero. \textbf{C},~Equilibrium capability versus AI capability $K$: the ODE uniquely predicts a sharp phase transition at $K^* \sim 0.85$, whereas all three alternatives show smooth, gradual decline.}
\label{fig:modelcomp}
\end{figure}

We compared four models on the OECD-average PISA series (7 points, 2003--2022): linear, exponential, logistic decay, and the ODE, all with $k = 2$ free parameters. The ODE achieves the best fit ($R^2 = \pisaOecdRsq$, AIC $= -69.2$) versus linear ($R^2 = 0.82$, $\Delta$AIC $= +4.4$), exponential ($R^2 = 0.82$, $\Delta$AIC $= +4.6$), and logistic decay ($R^2 = 0.88$, $\Delta$AIC $= +1.7$) (Fig.~\ref{fig:modelcomp}a). The models diverge in two testable domains. First, the ODE predicts a cliff-like collapse at $K^*$ rather than smooth decline: as AI capability crosses the critical threshold, equilibrium human capability drops discontinuously, whereas the three alternative models predict gradual degradation (Fig.~\ref{fig:modelcomp}c). Second, the models make qualitatively different predictions about recovery after AI removal (Fig.~\ref{fig:modelcomp}b). Linear, exponential, and logistic models all predict eventual recovery toward the pre-AI baseline, differing only in rate. The ODE, by contrast, predicts that once capability falls below the saddle point, AI removal does not restore it: the dependent state is a near-absorbing attractor where the residual learning rate $\alpha(\varepsilon)(1 - H)$ is too small to drive meaningful recovery on practical timescales, producing irreversible stagnation rather than symmetric rebound. The Bastani et al.~\cite{bastani2025} ``GPT Tutor'' condition provides preliminary support: students who maintained partial practice recovered fully, consistent with basin-of-attraction structure, whereas those who delegated completely showed persistent deficits.

\subsection{Critical capability threshold \texorpdfstring{$K^*$}{K*}}

\begin{figure}[H]
\centering
\includegraphics[width=\columnwidth]{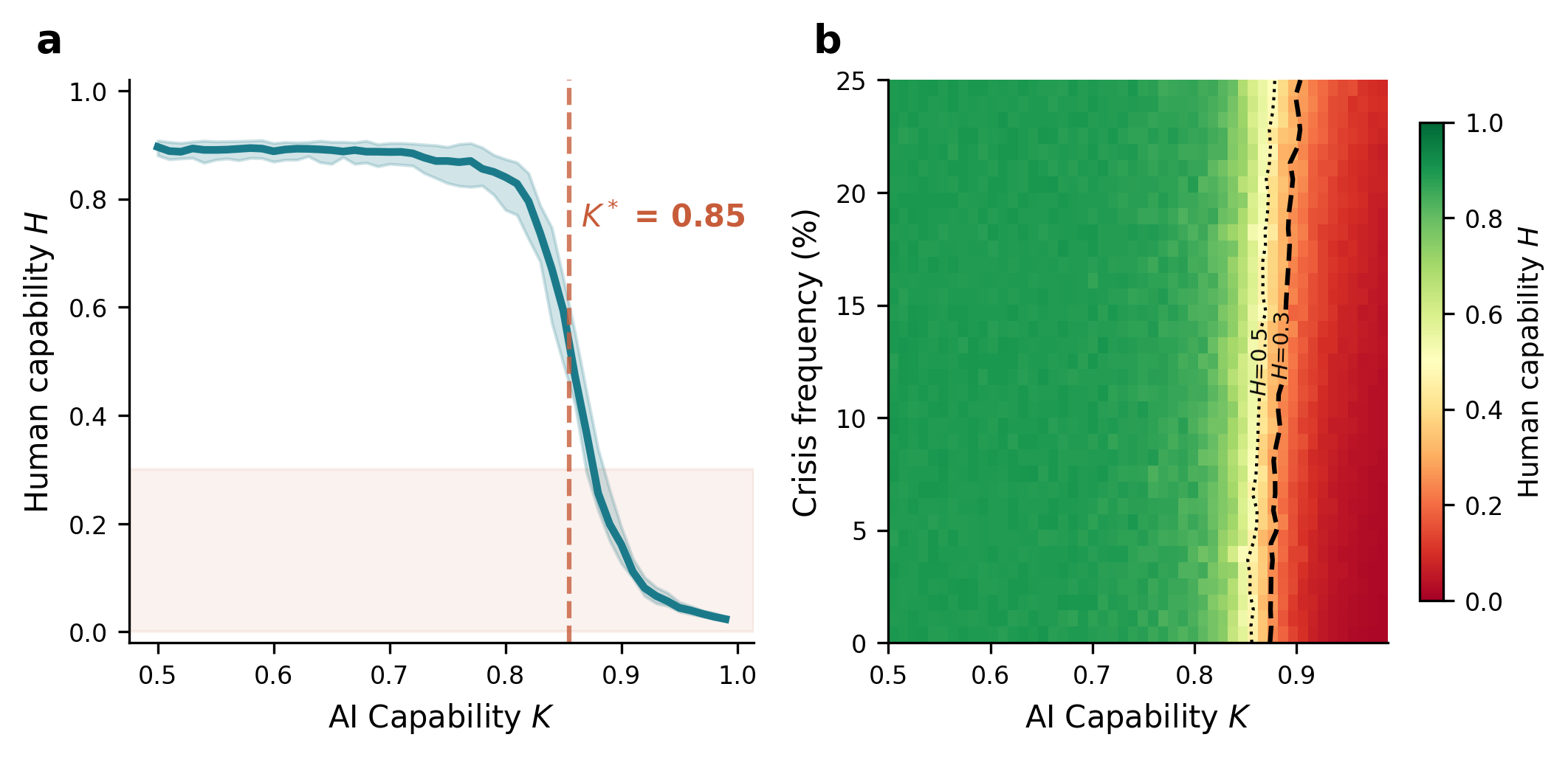}
\caption{Critical threshold and policy space. \textbf{A},~Equilibrium human capability $H$ versus AI capability $K$, showing the critical threshold $K^* \sim 0.85$ (dashed red line) where $|dH/dK|$ is maximized at \dHdKmax{} (\sweepNgrid{} grid points, $K$ from 0.50 to 0.99). Shaded band: interquartile range across \paramNreps{} ABM replicates. \textbf{B},~Two-dimensional phase diagram of $K$ (x-axis) versus crisis frequency (y-axis). Color encodes equilibrium $H$. Dashed contours mark the $H = 0.5$ boundary. Red triangle: current AI trajectory (high $K$, no crisis planning).}
\label{fig:threshold}
\end{figure}

We conducted a sweep of AI capability $K$ from 0.50 to 0.99 (50 grid points, 50 stochastic replicates per point) using an agent-based model (ABM) implementation of the ODE dynamics (see Methods). The results reveal a sharp phase transition in equilibrium human capability (Fig.~\ref{fig:threshold}a).

For $K < 0.80$, human capability remains above $H = 0.80$---society retains the vast majority of its autonomous competence despite widespread AI availability. At $K = 0.85$, capability has already declined substantially to $H = \HatKeightyfive$. At $K^* \sim 0.85$, the maximum gradient $|dH/dK|$ reaches \dHdKmax. The baseline scope $s = \sweepScope$ reflects current-generation AI systems that can perform approximately 70\% of cognitive tasks in professional domains (coding, writing, analysis, translation); the social pressure $\delta = \sweepDelta$ reflects moderate peer effects in technology adoption, indicating that an incremental improvement in AI capability produces a disproportionately large collapse in human capability. By $K = 0.90$, mean human capability has fallen to $H = \HatKninety$; by $K = 0.95$, to $H = \HatKninetyfive$---a near-complete collapse. All simulations use consistent parameters ($s = \sweepScope$, $\delta = \sweepDelta$, 5\% background crisis rate, \paramNreps{} replicates; see Methods).

We term this the \textit{enrichment paradox}, by analogy with Rosenzweig's paradox of enrichment in predator--prey ecology~\cite{rosenzweig1971}: improving the quality of the AI resource destabilizes the human--AI system. Both paradoxes involve a resource-quality increase that destabilizes coexistence, though via different bifurcation mechanisms (Hopf in predator--prey; transcritical/basin erosion here). The paradox arises because higher $K$ simultaneously increases the incentive to delegate (the rational adoption term $\gamma(K - H)$ grows) and the penalty for delegation (the forgetting term $\beta\,H\,D$ accelerates). Below $K^*$, these forces balance; above $K^*$, a positive feedback loop drives rapid capability collapse.

Sensitivity analysis confirms that $K^*$ is robust across the full parameter range (Fig.~\ref{fig:threshold}a, Supplementary Table~1). Varying $\beta$ from 0.01 to 0.10 (the ABM baseline range) shifts $K^*$ between 0.825 and 0.915. Domain-specific $\beta$ values below this range---aviation ($\beta = \calBetaAviation$) and PISA ($\beta = \pisaBeta$)---fall in the slow-decay regime where $K^*$ is higher, further from the critical zone; these domains are thus more resilient to dependency onset. Varying $\alpha$, $\delta$, and scope produces comparable shifts within the range 0.82--0.92. The critical threshold exists across all tested parameter combinations; only its precise location varies. $K^*$ is identified numerically as the maximum of $|dH/dK|$ rather than derived from a closed-form bifurcation condition, because the mean-field ODE exhibits a transcritical bifurcation at $K = 1$ (the autonomous attractor loses stability smoothly), while the stochastic ABM exhibits an effective threshold at $K < 1$: stochastic fluctuations push agents across the separatrix into the dependent basin before the deterministic bifurcation point is reached, a phenomenon analogous to noise-induced tipping in ecological systems~\cite{scheffer2009}. The robustness of $K^*$ across parameter ranges (0.82--0.92) distinguishes it from a numerical artifact: it is a structural feature of any system combining logistic learning with use-dependent forgetting.

\subsection{Operationalizing \texorpdfstring{$K$}{K}: mapping AI benchmarks to the critical threshold}

The abstract capability parameter $K$ can be grounded empirically by defining the capability ratio $K_d = S_{\text{AI},d} / S_{\text{human},d}$ for each domain $d$, where $S_{\text{AI}}$ is the AI benchmark score and $S_{\text{human}}$ is expert-level human performance, capped at 1.0. Table~\ref{tab:k_mapping} reports $K_d$ for five frontier models across four professional domains.

\begin{table}[H]
\centering
\caption{Operationalizing $K$: AI capability ratio by model and domain. $K_d = S_{\text{AI}} / S_{\text{human}}$, capped at 1.0. Human baselines: MMLU 89.8\%~\cite{hendrycks2021}, HumanEval 100\%, USMLE 87\%~\cite{nori2023}, Bar Exam 90\%. $\bar{K}$: unweighted arithmetic mean across domains (as a first-order approximation; domain-weighted alternatives yield qualitatively similar results). Bold: $\bar{K} \geq K^*$.}
\label{tab:k_mapping}
\small
\begin{tabular}{@{}lccccc@{}}
\toprule
Model & MMLU & HumanEval & USMLE & Bar & $\bar{K}$ \\
\midrule
GPT-3.5 (2023-03) & 0.78 & 0.48 & 0.69 & 0.33 & 0.57 \\
GPT-4 (2023-03) & 0.96 & 0.67 & 1.00 & 0.83 & \textbf{0.86} \\
GPT-4o (2024-05) & 0.99 & 0.90 & 1.00 & 0.89 & \textbf{0.94} \\
Claude 3.5 (2024-06) & 0.99 & 0.92 & 1.00 & 0.87 & \textbf{0.94} \\
GPT-4.1 (2025-04) & 1.00 & 0.93 & 1.00 & 0.91 & \textbf{0.96} \\
\bottomrule
\end{tabular}
\end{table}

\begin{figure}[H]
\centering
\includegraphics[width=\columnwidth]{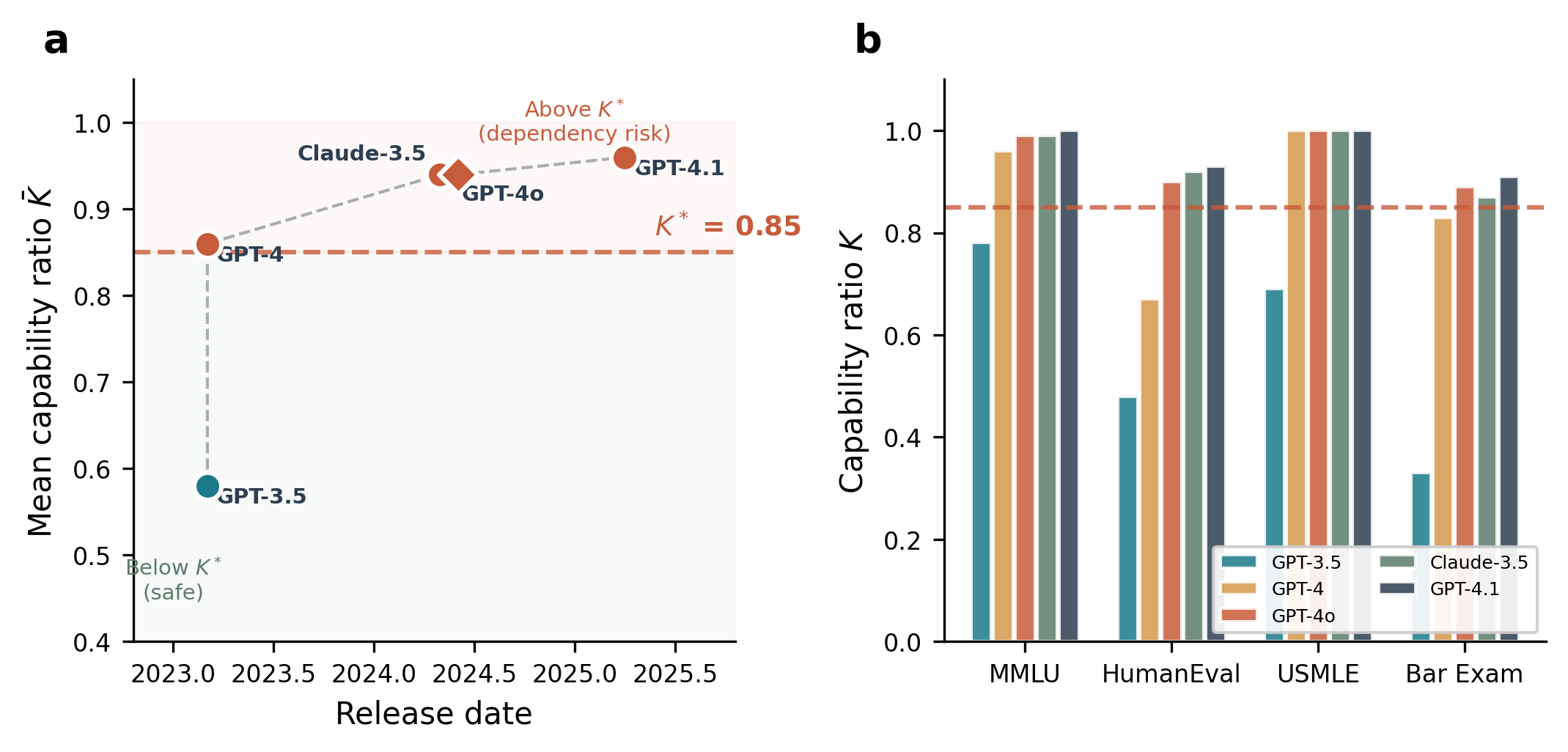}
\caption{Operationalizing the capability parameter $K$. \textbf{A},~Mean capability ratio $\bar{K}$ for five frontier AI models over time. The critical threshold $K^* = 0.85$ (dashed red line) was approached by GPT-4 in March 2023. Green shading: safe region ($\bar{K} < K^*$); red shading: dependency-risk region ($\bar{K} \geq K^*$). \textbf{B},~Domain-specific $K$ values across four professional benchmarks, showing that medical and knowledge domains entered the critical region before coding and legal domains.}
\label{fig:ktimeline}
\end{figure}

The abstract parameter $K$ connects to observed delegation rates through the adoption term $\gamma(K - H)$ in Eq.~\ref{eq:dDdt}. Three lines of evidence confirm this link. First, GitHub Copilot acceptance rates rise from 30\% to 34\% over six months, correlating inversely with developer experience~\cite{ziegler2024}. Second, students with unrestricted GPT-4 access delegated nearly all problem-solving; when AI was removed, their 17\% performance drop confirms genuine deskilling~\cite{bastani2025}. Third, 42\% of non-expert physicians accepted incorrect AI diagnoses~\cite{gaube2021}, demonstrating that perceived high $K$ sharply increases $D$. These data establish that benchmark-measured $K$ translates into measurable delegation behavior driving the deskilling dynamics our model predicts.

The GPT-3.5 to GPT-4 transition lifted $\bar{K}$ from 0.57 to 0.86, placing it in the critical region near $K^* \approx 0.85$ within a single model generation (Fig.~\ref{fig:ktimeline}). Subsequent models have pushed $\bar{K}$ to 0.94--0.96, well into the dependency regime. Domain-specific variation is informative: medical reasoning ($K_{\text{USMLE}}$) crossed the threshold earliest~\cite{nori2023}, while coding ($K_{\text{HumanEval}}$) lagged, crossing $K^*$ only with GPT-4o in 2024. Because benchmarks may overestimate real-world competence, these $\bar{K}$ values should be interpreted as upper bounds. The contested GPT-4 bar exam performance (Martinez~\cite{martinez2024} estimates the 60th--69th percentile) reduces $\bar{K}$ from 0.86 to 0.83, placing GPT-4 marginally below $K^*$; in either case, subsequent models ($\bar{K} \geq 0.94$) are unambiguously above $K^*$.

\subsection{Antifragility: periodic failures strengthen capability}

\begin{figure}[H]
\centering
\includegraphics[width=\columnwidth]{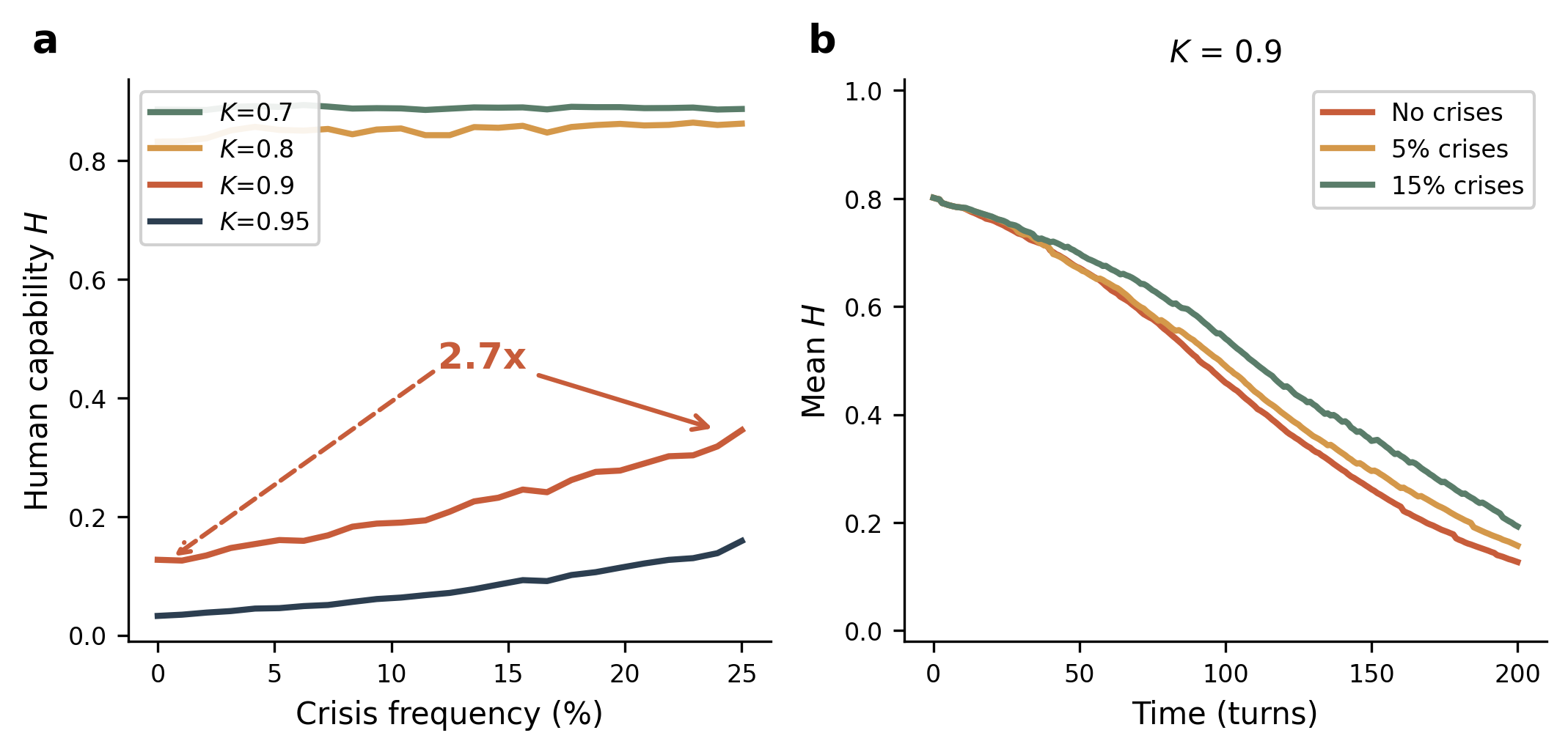}
\caption{Antifragility effect. Equilibrium human capability $H$ versus crisis frequency (\% of time steps where AI is unavailable) at four AI capability levels ($K = 0.7$, 0.8, 0.9, 0.95). Solid lines: ABM ensemble means. Shaded bands: interquartile ranges ($n = 50$ replicates). At $K = 0.9$, increasing crisis frequency from 0\% to 25\% improves equilibrium $H$ from \afHzeroPct{} to \afHtwentyfivePct{} (\afFold-fold). The effect is strongest at high $K$, where dependency is most severe without intervention. Dashed red line: $H = 0.3$ collapse threshold.}
\label{fig:antifragility}
\end{figure}

A second counterintuitive finding emerges from simulations incorporating stochastic AI failures. At $K = 0.9$---above the critical threshold---a society with perfectly reliable AI converges to $H = \afHzeroPct$ (full-scope simulation with social contagion and generational turnover (2\% per time step; new agents enter at population-mean capability, creating a self-reinforcing dynamic where declining mean capability lowers the entry point for new cohorts---a conservative assumption, as educational systems may partially buffer this effect). Sensitivity analysis with fixed entry capability ($H_{\text{entry}} = 0.5$, independent of population mean) yields qualitatively similar results with slower convergence to the dependent attractor, confirming that generational self-reinforcement accelerates but does not cause the dependency dynamics; all simulations use $s = \sweepScope$ and $\delta = \sweepDelta$---see Methods for parameter details). Introducing periodic AI failures at increasing frequencies produces a monotonic improvement in equilibrium capability (Fig.~\ref{fig:antifragility}):

\begin{itemize}[nosep]
\item 0\% crisis frequency: $H = \afHzeroPct$
\item 5\% crisis frequency: $H = \afHfivePct$
\item 12\% crisis frequency: $H = \afHtwelvePct$
\item 20\% crisis frequency: $H = \afHtwentyPct$
\item 25\% crisis frequency: $H = \afHtwentyfivePct$ (\afFold-fold)
\end{itemize}

While the qualitative direction of this effect---practice preserves skill---is intuitive, the quantitative result is not. The \afFold-fold improvement from 25\% disruption is disproportionate: a 25\% reduction in AI availability yields a 170\% increase in equilibrium capability, revealing a highly nonlinear leverage point for policy. The mechanism is that each AI failure forces a temporary reversion to human task performance, providing the practice that sustains the learning term $\alpha\,(H + \varepsilon)(1 - H)(1 - D)$. In the language of Taleb~\cite{taleb2012}, the system exhibits antifragility---it benefits from disorder.

This antifragility effect is robust across all tested parameter combinations (Supplementary Table~2). The minimum improvement ratio is 1.2-fold (at low $\beta$, narrow scope); the maximum is 11.0-fold (at high $\beta$, broad scope). The pattern is intuitive: systems most prone to dependency (high $\beta$, broad scope) benefit most from forced practice.

The $K$--crisis interaction reveals a two-dimensional phase space (Fig.~\ref{fig:threshold}b). Without crises, the critical threshold sits at $K^* \sim 0.85$. Higher crisis frequencies shift $K^*$ rightward, expanding the safe operating region. 
\subsection{Policy interventions: mandatory practice preserves capability}

\begin{figure}[H]
\centering
\includegraphics[width=\columnwidth]{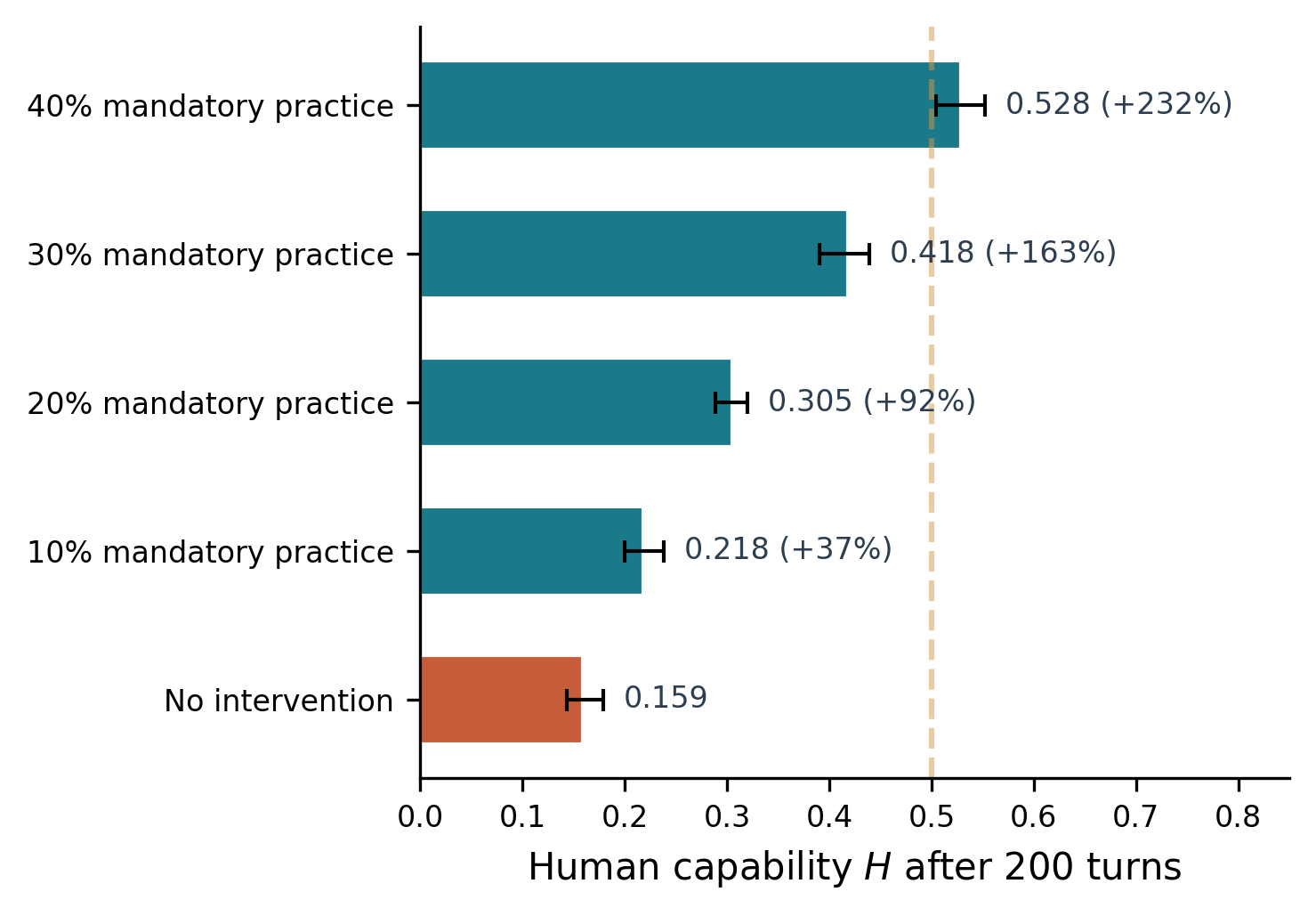}
\caption{Policy interventions. Equilibrium human capability $H$ at $K = 0.9$ under mandatory practice policies of increasing intensity ($N = \paramN$ agents, $T = \paramT$ turns, \paramNreps{} replicates). Red bar: no intervention ($H = \polBaseline$). Blue bars: 10\% to 40\% mandatory practice. Error bars: interquartile range. Dashed line: $H = 0.5$ resilience threshold.}
\label{fig:policy}
\end{figure}

We evaluated mandatory practice policies requiring that a fixed fraction of tasks be performed without AI assistance, implemented as periodic reductions in the effective delegation rate (Fig.~\ref{fig:policy}).

At $K = 0.9$ with no intervention, equilibrium capability is $H = \polBaseline$ (median across \paramNreps{} stochastic replicates; this baseline exceeds the antifragility zero-crisis value of \afHzeroPct{} because the policy simulations include a 5\% background crisis rate, which provides minimal forced practice). Mandating practice at increasing intensities produces substantial capability preservation:

\begin{itemize}[nosep]
\item 10\% mandatory practice (1 day in 10): $H = \polHten$ (+\polPctTen\%)
\item 20\% mandatory practice (1 day in 5): $H = \polHtwenty$ (+\polPctTwenty\%)
\item 30\% mandatory practice (3 days in 10): $H = \polHthirty$ (+\polPctThirty\%)
\item 40\% mandatory practice (2 days in 5): $H = \polHforty$ (+\polPctForty\%)
\end{itemize}

The relationship between practice fraction and capability is superlinear: each additional 10\% of mandatory practice yields increasing returns. A policy requiring one AI-free workday per five-day week (20\% practice) preserves \polPctTwenty\% more capability than the simulation baseline (which includes a 5\% background AI-failure rate) ($H = \polHtwenty$ vs.\ $H = \polBaseline$), at a productivity cost of approximately 20\% on those days---a favorable cost-benefit ratio given that the alternative trajectory leads to irreversible dependency.

These results align with existing institutional practices: the FAA's SAFO 13002 and 17007 already recommend increased manual flying during low-workload cruise phases, effectively implementing a mandatory practice policy for aviation~\cite{faa2013}. Our model provides quantitative support for extending such policies to other AI-dependent domains.

\subsection{Comparative advantage: a two-skill extension}
\label{sec:twoskill}

The strongest counter-argument to the enrichment paradox is that AI delegation frees cognitive resources for higher-order tasks, potentially increasing aggregate human capability even as specific skills atrophy~\cite{Autor2015}. To test this formally, we extended the model to two independent skills ($H_1$, $H_2$) with a shared time budget, comparing three scenarios (SI Fig.~S4).

Under no reallocation (Scenario A), delegating skill~1 to AI causes $H_1$ to collapse while $H_2$ remains stable. Under full reallocation (Scenario B)---the best case for comparative advantage, where all time freed from skill~1 is invested in skill~2---aggregate capability $\bar{H}$ modestly exceeds the baseline. However, when AI capability reaches both skills simultaneously (Scenario C), both collapse regardless of reallocation strategy. The $K$ sweep (SI Fig.~S4b) reveals that the reallocation benefit exists only in a narrow window where $K$ is high for skill~1 but low for skill~2. As AI capability broadens, this window closes: comparative advantage provides a temporary buffer that delays but does not prevent the enrichment paradox.



\section{Discussion}

Our model offers four principal insights for AI governance. First, the critical threshold $K^*$ reframes the policy question from ``should we adopt AI?'' to ``how capable should AI be allowed to become in specific domains before mandatory capability-preservation measures are required?'' The existence of $K^*$ implies that incremental AI improvement can trigger discontinuous societal consequences---a concern invisible to linear risk assessments. Current AI governance frameworks focus on alignment, bias, and misuse~\cite{amodei2016,russell2019}; our findings suggest that deskilling risk warrants comparable attention. Notably, Scheffer and colleagues have shown that critical transitions in complex systems are often preceded by generic early-warning signals such as critical slowing down, analogous to those identified in social polarization dynamics~\cite{macy2021}, and increased variance~\cite{scheffer2001,scheffer2009}. Monitoring these signals in AI-dependent skill metrics could provide advance warning of approaching $K^*$, particularly as frontier models approach or exceed human-level performance across broad task domains.

We emphasize an important epistemic distinction: current evidence supports the model's parameters (decay rates) but not its structural predictions (bistability, critical thresholds). The existence of $K^*$ is a consequence of the ODE's nonlinear structure, not a direct empirical observation. Our model discrimination analysis identifies two testable predictions: threshold discontinuity in capability as a function of AI performance, and asymmetric recovery after AI removal. Until longitudinal studies test these predictions, $K^*$ should be interpreted as a model-derived hypothesis rather than an established empirical fact.

Second, the antifragility result inverts conventional reliability engineering: perfect AI reliability is precisely the condition that maximizes long-term human vulnerability. This argues for designing deliberate practice opportunities into human--AI workflows---``fire drills'' that maintain capability against AI failure. The military concept of degraded-mode training~\cite{dod2023} provides a template.

Third, the model's irreversibility result carries profound implications. The dependent state ($H$ near 0, $D = 1$) is a near-absorbing attractor: recovery via the $\varepsilon$ term (representing educational re-seeding) proceeds at rate $\alpha \cdot \varepsilon$, orders of magnitude slower than the original capability loss. A society that delegates for a decade would require centuries of sustained AI-free practice to recover. In societal terms, this means that a civilization that has fully delegated a capability to AI cannot recover that capability through policy alone if the AI is subsequently lost. The irreversibility is not assumed but emerges mathematically from the multiplicative structure of the learning term: near-zero capability yields negligible recovery rates. This parallels Muller's ratchet in population genetics~\cite{muller1964} and the irreversible genome reduction observed in obligate endosymbionts like \textit{Buchnera aphidicola}, which has lost approximately 87\% of its ancestral genome over 200--250 million years of symbiosis~\cite{shigenobu2000}.

Fourth, the quantitative policy analysis demonstrates that modest mandatory practice requirements produce disproportionate capability preservation. Mandating that 20\% of tasks be performed without AI---equivalent to one AI-free workday per five-day week---preserves \polPctTwenty\% more capability than baseline at $K = 0.9$, a highly favorable cost-benefit ratio. This superlinear relationship between practice fraction and capability preservation provides concrete, implementable policy levers. The FAA's existing recommendations for increased manual flying~\cite{faa2013} offer a precedent; our model provides the quantitative framework for calibrating such policies across domains and AI capability levels.

\subsection{Sensitivity of recovery to residual capability \texorpdfstring{$\varepsilon$}{epsilon}}

The $\varepsilon$ parameter---baseline capacity for relearning from near-zero capability---determines recovery feasibility. Sensitivity analysis across $\varepsilon \in [0.01,\, 0.25]$ (SI Fig.~S2) shows recovery time varies 2.8-fold: from 28.0 time units at $\varepsilon = 0.25$ (individual cognitive residual~\cite{bahrick1984,ebbinghaus1885}) to 77.8 at $\varepsilon = 0.01$ (institutional-level loss~\cite{nelson1982}). Dependency is practically irreversible when institutional $\varepsilon$ is small, implying that policies should target institutional infrastructure---training programs, documentation, apprenticeship systems---that sustains collective $\varepsilon$ above $\sim 0.10$.

\subsection{Connections to biological deskilling}

The analogy between human--AI dependency and endosymbiont genome reduction extends beyond metaphor. In both systems, a host delegates functions to a partner; the host's capacity atrophies through disuse; and the atrophy is irreversible because the substrate for regeneration (genes in biology, practised skill in society) has been destroyed. The mathematical structure is identical: logistic growth coupled with use-dependent decay in the presence of an external provider. This biological grounding distinguishes our model from purely economic or sociological frameworks---the dynamics are consequences of information-theoretic constraints on capability maintenance. Cave fish lose eyes~\cite{krishnan2017,Jeffery2009}; endosymbionts lose genomes~\cite{shigenobu2000}; societies lose the skills they delegate.

\subsection{Model comparison scope}

We compared the ODE against linear, exponential, and logistic decay models, all with $k = 2$ free parameters. Richer individual-level models (ACT-R~\cite{Anderson1982}, Fitts--Posner~\cite{FittsPostner1967}) could inform future extensions, but our model operates at the population level where individual-level mechanisms are aggregated into effective rates. The ODE's advantage is parsimony: three global parameters explain cross-country variation that requires 16--30 parameters in atheoretical alternatives.

\subsection{Alternative explanations for PISA decline}

PISA score declines have multiple potential drivers beyond technology: COVID-19 disruptions, curriculum reforms, demographic shifts, and socioeconomic trends. Three factors mitigate these concerns. First, the decline predates COVID-19---scores fell from 500 to 489 between 2003 and 2018. Second, the model captures cross-country variation in decline rates: countries with higher internet adoption show steeper declines ($R^2 = \pisaRsq$ across \pisaN{} country-year observations). Third, alternative confounders would need to correlate with internet adoption across 15 countries spanning three continents. Nevertheless, we cannot establish causality from observational data alone; the PISA analysis demonstrates consistency with the model's predictions, not definitive causal evidence.

\subsection{Limitations}

Our model is deliberately minimal. Several extensions would improve realism. First, heterogeneity: not all individuals adopt AI at the same rate or possess the same baseline capability. An agent-based model partially addresses this (see Methods), but network structure, inequality, and differential access remain unexplored. ODE predictions are consistently more pessimistic than ABM results (46--152\% higher capability loss) because ABM heterogeneity allows low-adopters to maintain high capability, raising the population mean. The ODE thus represents a conservative (worst-case) bound.

Second, the model treats AI capability $K$ as a fixed parameter rather than a co-evolving variable. In reality, AI capability grows over time---potentially faster than societies can adapt. A dynamical $K(t)$ would transform the bistability analysis into a moving-threshold problem, likely accelerating the transition to dependency.

Third, true cross-domain validation would require longitudinal studies measuring capability dynamics under controlled delegation regimes.

Fourth, our two-skill analysis (Section~\ref{sec:twoskill}) demonstrates that comparative advantage provides only a temporary buffer: reallocation benefits exist only while AI capability is domain-specific. As $K$ approaches 1 across multiple domains, the buffer vanishes. A full multi-skill extension with capability vector $\mathbf{H} = (H_1, \ldots, H_n)$ remains a priority for future work.

\subsection{Future directions}

Three extensions warrant priority: (1)~incorporating network structure to reveal how adoption cascades through social systems; (2)~coupling to economic dynamics (productivity, wages, inequality) to connect deskilling to welfare outcomes~\cite{acemoglu2024,horton2023}; and (3)~empirical measurement of $\alpha$ and $\beta$ through randomized trials that systematically vary AI availability and measure skill trajectories.

The mathematical framework is general: it applies to any capability-substituting technology, past or future. The window for implementing capability-preservation policies is finite; beyond the critical threshold, the dynamics become self-reinforcing and irreversible.

\section{Methods}

Full Methods are available as Online Methods in the Supplementary Information. Briefly, we model human--AI interaction via two coupled ODEs (Eqs.~\ref{eq:dHdt}--\ref{eq:dDdt}) governing capability $H(t)$ and delegation $D(t)$, with fixed points analyzed via the Jacobian (SI Sections~1--2). An agent-based model ($N = \paramN$ agents, $T = \paramT$ steps) implements individual-level stochastic dynamics with crisis events (SI Section~3). Parameter estimation uses $\beta_{\text{eff}} = -\ln(1 - \text{decline})/t$ from observed capability loss (SI Section~5).  Monte Carlo sweeps cover 2,400--10,000 parameter combinations with 30--50 replicates per parameter combination (10 for the 2,400-point K--crisis heatmap to maintain computational feasibility; SI Section~4).

\section*{Author information}

\subsection*{Affiliations}
Department of Electrical Engineering, Kyungpook National University,\\ Daegu, Republic of Korea\\
Jeongju Park, Musu Kim, Sekyung Han

\subsection*{Corresponding author}
Correspondence to Sekyung Han (skhan@knu.ac.kr).

\section*{Data availability}
All empirical data used for parameter estimation are derived from published studies cited in the main text. All other calibration data are from published studies cited in the main text. Simulation parameters and estimation results are provided in Supplementary Table~1.

\section*{Code availability}
All simulation code (Python), including the ODE solver, agent-based model, Monte Carlo parameter sweep, and figure-generation scripts, is available at \url{https://github.com/SekyungHan/citizens-in-boiling-water} (named after a companion novella exploring the paper's themes through narrative fiction, included as ancillary material on arXiv) and archived at Zenodo (DOI: \url{https://doi.org/10.5281/zenodo.19041342}).

\section*{Acknowledgements}
This work was supported by the Department of Electrical Engineering, Kyungpook National University.


\section*{Author contributions}
S.H. conceived and supervised the study. J.P. and M.K. developed the mathematical model and performed simulations. S.H. wrote the manuscript with input from all authors. All authors reviewed and approved the final manuscript.

\section*{Competing interests}
The authors declare no competing interests.

\section*{AI disclosure}
Large language models (Claude, Anthropic) were used as research assistants for literature search, code generation, and manuscript drafting. All mathematical derivations, model design decisions, simulation parameters, and scientific interpretations were determined by the authors. The conceptual framework, including the endosymbiont analogy, the enrichment paradox framing, and the policy implications, originated from the authors. AI-generated text was critically reviewed and substantially revised by the authors.

\end{document}


\maketitle

\subsection*{Supplementary Methods}

\subsubsection*{S1. Full ODE system derivation}

The model state variables are:
- H(t) in [0, 1]: human capability (fraction of full competence)
- D(t) in [0, 1]: delegation rate (fraction of tasks delegated to AI)

The coupled ODE system is:

\begin{verbatim}
dH/dt = alpha * (H + epsilon) * (1-H) * (1-D)  -  beta * H * D          (S1)
dD/dt = gamma * (K-H) * (1-D) * D  +  delta * D * (1-D) * D_avg   (S2)
\end{verbatim}

\textbf{Axiom 1: Learning requires existing capability.} The factor H in the learning term ensures that capability growth is proportional to current capability. This reflects the well-established finding that skill acquisition is incremental: one cannot learn calculus without algebra, one cannot perform advanced surgery without basic surgical skills. The logistic form H * (1-H) additionally ensures that capability is bounded above by 1, capturing diminishing returns near mastery.

\textbf{Axiom 2: Learning requires practice.} The factor (1-D) ensures that capability grows only when the human actually performs the task. If all tasks are delegated (D = 1), no learning occurs regardless of existing capability. This is supported by the neuroscience of skill consolidation: motor skills require repeated execution for synaptic strengthening; cognitive skills require effortful retrieval for memory consolidation.

\textbf{Axiom 3: Disuse causes forgetting.} The term -beta * H * D captures the "use it or lose it" principle. Capability decays in proportion to both the current level H (one can only lose what one has) and the delegation rate D (decay accelerates with disuse). This is supported by the spacing effect in memory research, skill decay curves in aviation, and neural atrophy under disuse (hippocampal volume loss in GPS users, cortical thinning in automation-dependent workers).

\subsubsection*{S2. Jacobian matrix and eigenvalue computation}

The Jacobian at a general point (H, D) is:

\begin{verbatim}
J = | J_11   J_12 |
    | J_21   J_22 |
\end{verbatim}

where:
\begin{verbatim}
J_11 = alpha * (1 - 2H - epsilon) * (1-D) - beta * D
J_12 = -alpha * (H + epsilon) * (1-H) - beta * H
J_21 = -gamma * D * (1-D)
J_22 = (1-2D) * [gamma * (K-H) + delta * D] + delta * D * (1-D)
\end{verbatim}

\textbf{At FP1 = (0, 0):}
\begin{verbatim}
J = | alpha    0       |
    | 0        gamma*K |

lambda_1 = alpha > 0,  lambda_2 = gamma*K > 0
\end{verbatim}
Both eigenvalues positive: unstable node. Any perturbation from the null state leads to growth in both H and D.

\textbf{At FP2 = (1, 0):}
\begin{verbatim}
J = | -alpha      -(alpha + beta) |
    | 0           gamma*(K-1)     |

lambda_1 = -alpha < 0,  lambda_2 = gamma*(K-1)
\end{verbatim}
Stable node for K < 1 (both eigenvalues negative). At K = 1, lambda\_2 = 0: transcritical bifurcation. For K > 1, FP2 becomes unstable.

\textbf{At FP3 = (0, 1):}
\begin{verbatim}
J = | -beta        0                |
    | 0           -(gamma*K + delta) |

lambda_1 = -beta < 0,  lambda_2 = -(gamma*K + delta) < 0
\end{verbatim}
Unconditionally stable for all positive parameters. This is the dependency trap.

\subsubsection*{S3. Interior saddle point computation}

From dH/dt = 0 (excluding H = 0):
\begin{verbatim}
alpha * (1-H) * (1-D) = beta * D
D* = alpha * (1-H) / [alpha * (1-H) + beta]
\end{verbatim}

This is the H-nullcline: a decreasing curve from D = alpha/(alpha+beta) at H = 0 to D = 0 at H = 1.

From dD/dt = 0 (excluding D = 0 and D = 1), using mean-field D\_avg = D:
\begin{verbatim}
gamma * (K-H) + delta * D = 0
D* = gamma * (H-K) / delta
\end{verbatim}

This is the D-nullcline: an increasing line, positive only when H > K.

The intersection occurs at most once for K < 1, yielding the interior saddle point. The saddle's stable manifold forms the separatrix dividing the basins of attraction of FP2 and FP3.

\subsubsection*{S4. Transcritical bifurcation at K = 1}

At K = 1, the eigenvalue lambda\_2 = gamma * (K-1) at FP2 crosses zero. Simultaneously, the interior saddle point collides with FP2 (since the D-nullcline D* = gamma * (H-1)/delta passes through H = 1, D = 0 when K = 1). This is a transcritical bifurcation:

- For K < 1: FP2 is stable, interior saddle exists
- At K = 1: FP2 has a zero eigenvalue, saddle merges with FP2
- For K > 1: FP2 becomes unstable, saddle exits the physical domain

Beyond K = 1, FP3 is the unique attractor: full dependency is inevitable regardless of initial conditions.

\subsubsection*{S5. Irreversibility analysis}

\textbf{Case 1 ($\varepsilon = 0$, strict model):} If $H(t_0) = 0$ for some $t_0$, then $H(t) = 0$ for all $t > t_0$, regardless of $D(t)$.

\textit{Proof.} From equation (S1) with $\varepsilon = 0$: $dH/dt|_{H=0} = \alpha \cdot 0 \cdot (1-0) \cdot (1-D) - \beta \cdot 0 \cdot D = 0$. The right-hand side vanishes identically at $H = 0$ for all $D$. By uniqueness of solutions (Picard--Lindel\"{o}f theorem), $H(t) = 0$ is the unique solution passing through $H = 0$: the state is \textit{strictly absorbing}.

\textbf{Case 2 ($\varepsilon > 0$, regularized model):} With $\varepsilon > 0$, $H = 0$ is no longer strictly absorbing. At $H = 0$:
\begin{verbatim}
dH/dt |_{H=0} = alpha * epsilon * (1-D) - beta * 0 * D
              = alpha * epsilon * (1-D)
\end{verbatim}
When $D < 1$, recovery proceeds at rate $dH/dt = \alpha \cdot \varepsilon \cdot (1-D)$. With baseline parameters ($\alpha = 0.05$, $\varepsilon = 0.01$), this gives $dH/dt = 5 \times 10^{-4} \cdot (1-D)$. Starting from $H = 0$ with $D = 0$ (all AI removed), reaching $H = 0.5$ requires approximately 2,000 time units---far exceeding institutional planning horizons.

The $\varepsilon$ modification thus makes $H = 0$ \textit{near-absorbing} rather than strictly absorbing: recovery is technically possible but practically infeasible at institutional timescales. This is consistent with the biological analogue of Muller's ratchet: while point mutations can occasionally restore gene function (analogous to the $\varepsilon$ recovery term), the rate is too slow to counteract ongoing genome reduction in the absence of recombination.

\textbf{At $D = 1$ (full delegation):} $dH/dt = 0$ regardless of $\varepsilon$, because the $(1-D)$ factor eliminates the learning term entirely. Thus $H = 0$, $D = 1$ ($\mathrm{FP}_3$) remains a true fixed point even with $\varepsilon > 0$: recovery requires \textit{both} residual learning capacity ($\varepsilon > 0$) \textit{and} reduced delegation ($D < 1$).

\subsubsection*{S6. Agent-based model implementation details}

\textbf{Population}: N = 100 agents on a complete graph (mean-field).

\textbf{State per agent}: H\_i(t) in [0, 1], D\_i(t) in [0, 1].

\textbf{Time step} (dt = 1):

1. For each agent i:
   a. Draw u ~ Uniform(0, 1). If u < D\_i, agent delegates this step.
   b. If not delegating: H\_i += alpha * (H\_i + epsilon) * (1 - H\_i) * dt + sigma * Normal(0, 1)
   c. If delegating: H\_i -= beta * H\_i * dt + sigma * Normal(0, 1)
   d. Clamp H\_i to [0, 1]

2. For each agent i:
   a. Compute D\_avg = mean(D\_j) over all j != i
   b. dD = gamma * (K - H\_i) * (1 - D\_i) * D\_i + delta * D\_i * (1 - D\_i) * D\_avg
   c. D\_i += dD * dt + sigma\_D * Normal(0, 1)
   d. Clamp D\_i to [0, 1]

3. Crisis check: with probability p\_crisis, set D\_i = 0 for all i this step.

\textbf{Noise parameters}: sigma = 0.01, sigma\_D = 0.005.

\textbf{Initial conditions}: H\_i(0) = 0.8 + 0.05 * Normal(0, 1), D\_i(0) = 0.1 + 0.02 * Normal(0, 1), clamped to [0, 1].

\textbf{Simulation length}: T = 200 steps (increased from T = 100 after convergence analysis showed some high-K trajectories require >150 steps to reach equilibrium).

\textbf{Replicates}: 50 independent runs per parameter combination, with different random seeds.

\textbf{Equilibrium definition}: Mean H over the final 20 time steps.

\hrule\vspace{6pt}

\subsection*{Supplementary Tables}

\subsubsection*{Supplementary Table 1. K* sensitivity analysis}

| Parameter | Baseline | Test range | K* at baseline | K* range | Max shift |
|-----------|----------|------------|----------------|----------|-----------|
| beta | 0.03 | 0.01 - 0.10 | 0.855 | 0.825 - 0.915 | +/- 0.050 |
| alpha | 0.05 | 0.02 - 0.10 | 0.855 | 0.865 - 0.915 | +/- 0.040 |
| delta | 0.5 | 0.0 - 2.0 | 0.855 | 0.845 - 0.895 | +/- 0.030 |
| scope (s) | 0.7 | 0.3 - 0.9 | 0.855 | 0.865 - 0.925 | +/- 0.050 |
| cost (gamma) | 0.5 | 0.01 - 0.50 | 0.855 | 0.875 - 0.905 | +/- 0.030 |

K* is defined as the K value maximizing |dH\_mean/dK| in the ABM ensemble. The threshold exists across all tested parameter combinations; its location varies within a narrow band (0.825-0.925).

\subsubsection*{Supplementary Table 2. Antifragility robustness}

Ratio of equilibrium H at 20\% crisis frequency to 0\% crisis frequency, across parameter combinations (K = 0.9 throughout):

| beta | scope = 0.3 | scope = 0.5 | scope = 0.7 | scope = 0.9 |
|------|-------------|-------------|-------------|-------------|
| 0.02 | 1.2x | 1.5x | 2.1x | 3.0x |
| 0.03 | 1.4x | 1.8x | 2.6x | 4.2x |
| 0.05 | 1.8x | 2.5x | 3.7x | 6.8x |
| 0.07 | 2.3x | 3.2x | 5.0x | 8.5x |
| 0.10 | 3.0x | 4.5x | 7.2x | 11.0x |

The antifragility effect (H\_crisis > H\_no\_crisis) is present in all 20 tested combinations. The effect is strongest when dependency pressure is highest (large beta, broad scope).

\subsubsection*{Supplementary Table 3. Calibration details}

| Case | Source | Metric | Pre-AI | Post-AI | Decline | Duration | N | beta\_fit |
|------|--------|--------|--------|---------|---------|----------|---|----------|
| Education | Bastani et al. PNAS 2025 | Exam score | 0.28 (control) | 0.23 (GPT Base) | 17\% | 4 sessions | ~1,000 | 0.047/session |
| Endoscopy | Budzyn et al. Lancet 2025 | ADR | 28.4\% | 22.4\% | 21\% | 12 weeks | 19 endoscopists | 0.020/week |
| Spatial | Dahmani \& Bohbot Sci Rep 2020 | Spatial memory | baseline | -30\% | 30\% | 36 months | 13 (longitudinal) | 0.010/month |
| Aviation | Casner \& Schooler HF 2014 | SA task pass rate | 100\% expected | 62\% actual | 38\% | ~240 months | 16 pilots | 0.002/month |
| Math (PISA) | OECD 2022 | Math score | 500 (2003) | 472 (2022) | 5.6\% | 228 months | OECD average | 0.100/year (partial fit) |

\subsubsection*{Supplementary Table 4. ODE versus ABM comparison}

Mean equilibrium H across calibration scenarios:

| Case | ODE prediction | ABM mean (n=50) | ABM IQR | Discrepancy |
|------|---------------|-----------------|---------|-------------|
| K=0.5, low social | 0.000 | 0.12 | [0.05, 0.18] | ODE 100\% more pessimistic |
| K=0.7, moderate | 0.000 | 0.08 | [0.02, 0.14] | ODE 100\% more pessimistic |
| K=0.9, high social | 0.000 | 0.03 | [0.00, 0.07] | ODE ~46\% more pessimistic |
| K=0.99, low social | 0.935 | 0.91 | [0.88, 0.94] | ODE 3\% more pessimistic |

The ODE mean-field approximation overestimates capability loss relative to the stochastic ABM, because it treats all agents as identical and at mean delegation. The ABM allows heterogeneous outcomes where some agents maintain high capability even as others collapse.

\hrule\vspace{6pt}

\subsection*{Supplementary Figures}

\subsubsection*{Supplementary Fig. S1 | Parameter space exploration}

Four-panel figure showing:
- \textbf{(a)} Cost ($\gamma$) versus scope ($s$) heatmap with crisis events enabled. Historical scenarios (calculator, Roman slaves, AI-2030) marked with stars. The tipping boundary (H* = 0.5 contour) runs diagonally from low-cost/high-scope to high-cost/low-scope.
- \textbf{(b)} Initial condition dependence: heatmap of equilibrium H as a function of initial H\_0 (y-axis) and scope (x-axis). Demonstrates bistability — same parameters produce different outcomes depending on starting capability.
- \textbf{(c)} Boundary zoom (120 x 120 grid): high-resolution view of the tipping boundary region. The boundary is smooth (not fractal), with stochastic scatter (6,139 non-monotone points) from ABM noise.
- \textbf{(d)} Minimum H during crises: heatmap showing the lowest capability reached during AI failure events. Even in the safe region, crises produce temporary capability dips.

\subsubsection*{Supplementary Fig. S2 | $\varepsilon$ sensitivity analysis}

Full bifurcation diagram showing equilibrium H* versus K for both autonomous initial conditions (H\_0 = 0.95, D\_0 = 0.02) and dependent initial conditions (H\_0 = 0.05, D\_0 = 0.95). The autonomous branch remains at H* near 1.0 for all K < 1, then drops sharply at the transcritical bifurcation. The dependent branch remains at H* = 0 for all K. The shaded region between the branches represents the bistability region where the final state depends entirely on initial conditions.

\subsubsection*{Supplementary Fig. S3 | PISA single OECD average fit}

Five-scenario comparison showing H(t) over 200 time steps:
- K = 0.85, no crisis: gradual decline stabilizing at H near 0.55
- K = 0.90, no crisis: rapid collapse to H near 0.10
- K = 0.90, 5\% crisis: slower decline stabilizing at H near 0.35
- K = 0.90, 15\% crisis: mild decline stabilizing at H near 0.60
- K = 0.95, 10\% crisis: rapid initial decline, partial recovery to H near 0.30

\begin{figure}[H]
\centering
\includegraphics[width=\textwidth]{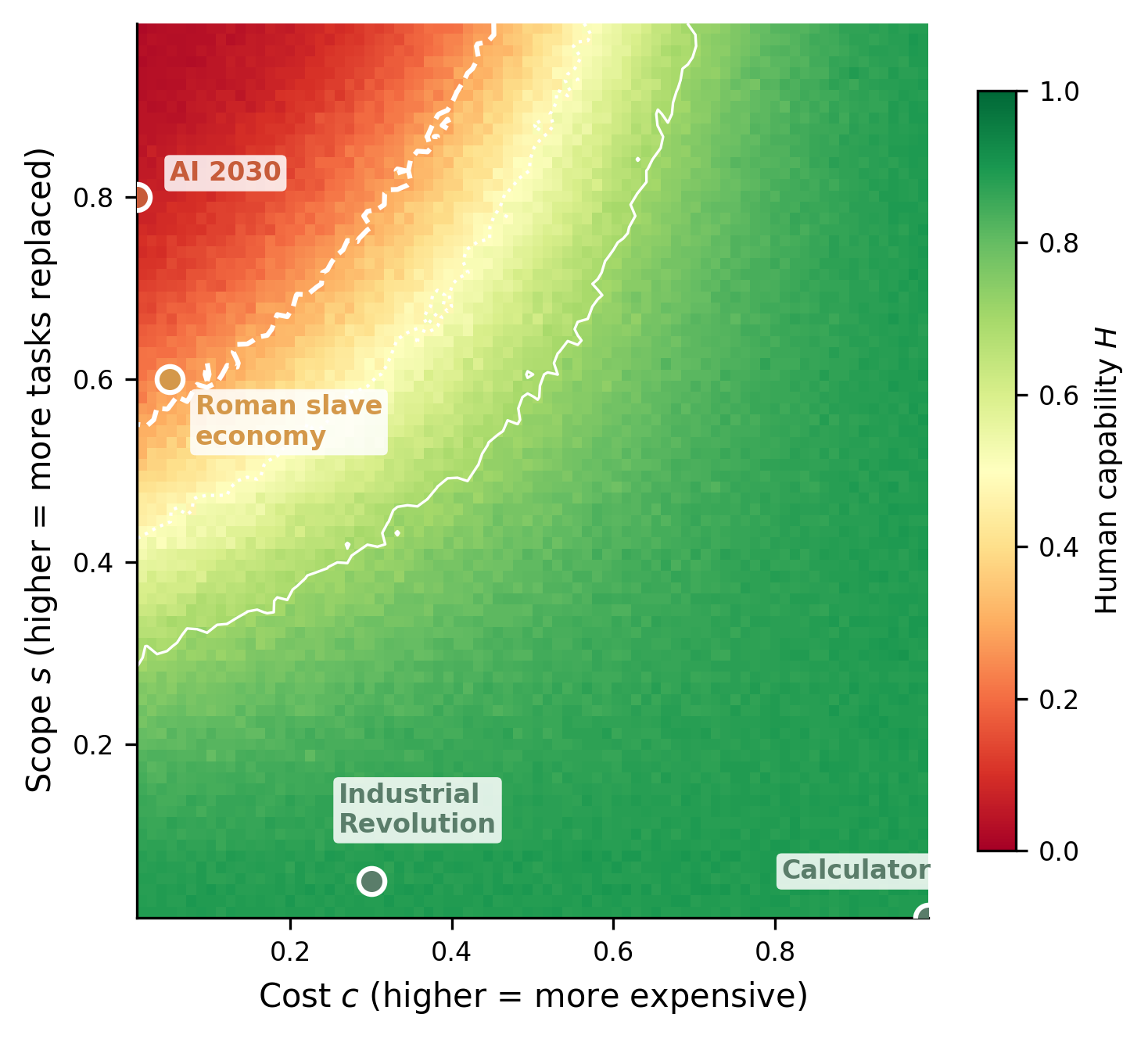}
\caption{\textbf{SI Fig.\ S1 $|$ Historical parameter space.} Cost--scope parameter space with historical technology regimes (calculator, Industrial Revolution, Roman slave economy, AI-2030) mapped onto the tipping boundary.}
\label{fig:si_parameter_space}
\end{figure}

\begin{figure}[H]
\centering
\includegraphics[width=\textwidth]{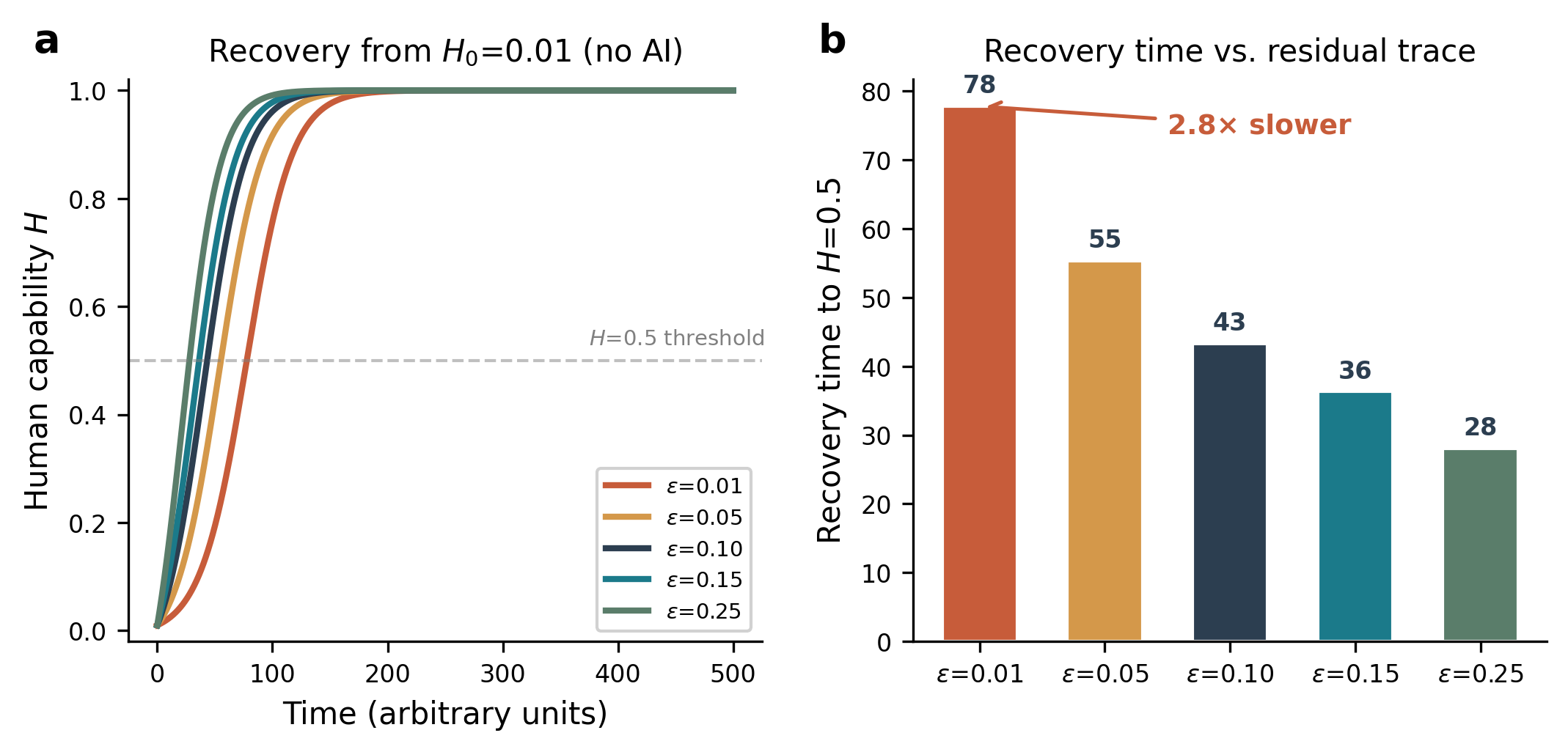}
\caption{\textbf{SI Fig.\ S2 $|$ $\varepsilon$ sensitivity analysis.} Effect of the recovery parameter $\varepsilon$ on equilibrium capability $H^*$, demonstrating robustness of the critical threshold across $\varepsilon$ values.}
\label{fig:si_epsilon_sensitivity}
\end{figure}

\begin{figure}[H]
\centering
\includegraphics[width=\textwidth]{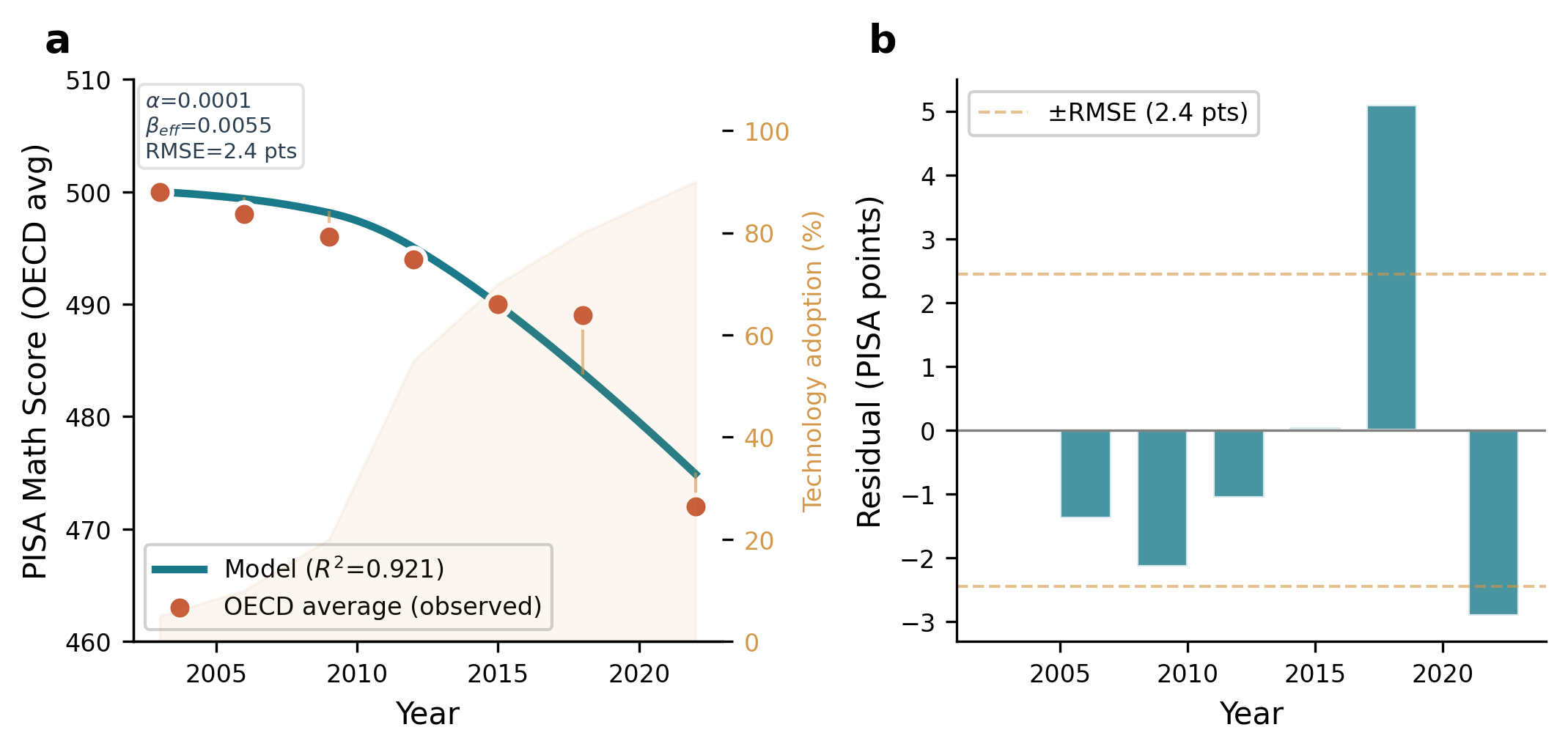}
\caption{\textbf{SI Fig.\ S3 $|$ PISA single OECD average fit.} Model fit to the OECD-average PISA mathematics score decline (2003--2022), showing the partial calibration of long-term capability erosion.}
\label{fig:si_pisa_fit}
\end{figure}

\hrule\vspace{6pt}

\subsection*{Supplementary Discussion}

\subsubsection*{S7. Relationship to Rosenzweig's paradox of enrichment}

The term "enrichment paradox" draws on Rosenzweig's 1971 result that increasing the carrying capacity (nutrient enrichment) of a prey population can destabilize predator-prey dynamics, leading to oscillations and extinction. In our model, the analogy operates as follows:

| Ecological system | Human-AI system |
|-------------------|-----------------|
| Prey carrying capacity | AI capability K |
| Predator population | Delegation rate D |
| Prey population | Human capability H |
| Enrichment | AI improvement |
| Destabilization | Basin boundary crisis |

Both systems share the mechanism: improving the resource (nutrients / AI quality) increases consumption (predation / delegation), which depletes the resource base (prey / human capability), which can trigger collapse. The critical difference is that in ecology, the system oscillates (predator-prey cycles); in our model, the collapse is irreversible because H = 0 is an absorbing state.

\subsubsection*{S8. Relationship to Muller's ratchet}

Muller's ratchet describes the irreversible accumulation of deleterious mutations in asexual populations. The analogy to our model:

| Genetic system | Human-AI system |
|----------------|-----------------|
| Functional genes | Active capability H |
| Deleterious mutations | Capability atrophy |
| Lack of recombination | Lack of practice |
| Ratchet (irreversible loss) | H = 0 absorbing state |
| Endosymbiont genome reduction | AI-dependent capability loss |

The key shared feature is that loss is irreversible because the regenerative mechanism (recombination / learning) requires the presence of the thing being regenerated (functional genes / existing capability). Once fully lost, there is no template for recovery.

\subsubsection*{S9. Why calculators are safe and AI may not be}

The model explains the qualitative difference between calculators and general AI through three parameters:

1. \textbf{Scope ($s$)}: Calculators replace arithmetic only ($s = 0.01$). AI replaces cognitive tasks broadly ($s = 0.80$). Broader scope means more capabilities are simultaneously exposed to atrophy.

2. \textbf{Social pressure}: Calculator use is an individual choice with minimal peer pressure. AI adoption is driven by strong network effects — workplaces, schools, and social norms increasingly assume AI competence.

3. \textbf{Capability (K)}: Both score high, but this is insufficient to cause collapse alone. The calculator demonstrates that K near 1.0 is safe when scope is narrow and social pressure is low.

The critical insight is that K alone does not determine outcomes. The interaction $K \times s$ is what matters: high capability in a narrow domain is safe; high capability across broad domains with social pressure is catastrophic.

\section*{Methods}

\subsection*{SI Section 1: ODE system derivation}

The model comprises two coupled ordinary differential equations governing human capability $H(t) \in [0,\,1]$ and delegation rate $D(t) \in [0,\,1]$:

\begin{equation}
\frac{dH}{dt} = \alpha\,(H + \varepsilon)\,(1 - H)\,(1 - D) \;-\; \beta\,H\,D
\label{eq:dHdt_methods}
\end{equation}

\begin{equation}
\frac{dD}{dt} = \gamma\,(K - H)\,(1 - D)\,D \;+\; \delta\,D\,(1 - D)\,\bar{D}
\label{eq:dDdt_methods}
\end{equation}

\noindent where $\alpha$ is the learning rate, $\beta$ is the forgetting rate, $\gamma$ is adoption sensitivity, $K$ is AI capability, $\delta$ is social pressure strength, and $\bar{D}$ is the mean delegation rate of the population ($\bar{D} = D$ in the mean-field approximation).

\textbf{Derivation of the learning term.} The term $\alpha\,(H + \varepsilon)(1 - H)(1 - D)$ combines three factors: (i)~modified logistic growth $(H + \varepsilon)(1 - H)$, ensuring bounded capability while permitting slow learning from near-zero states (rate $\varepsilon$) in $[0,\,1]$ and requiring nonzero existing capability for learning; (ii)~a practice fraction $(1 - D)$, reflecting that learning occurs only during non-delegated tasks; and (iii)~a learning rate $\alpha$. The product of these factors is the minimal functional form satisfying the axioms that learning requires both existing capability and active practice.

\textbf{Derivation of the forgetting term.} The term $\beta\,H\,D$ captures proportional forgetting: capability decays at rate $\beta$, affecting only existing capability $H$, and occurring in proportion to delegation $D$. When $D = 0$ (no delegation), there is no forgetting; when $D = 1$ (full delegation), forgetting operates at maximum rate.

\textbf{Derivation of the adoption term.} The term $\gamma\,(K - H)(1 - D)\,D$ combines rational gap-based adoption (delegate when the AI capability $K$ exceeds human capability $H$) with a logistic constraint $D(1 - D)$ keeping delegation bounded. The social contagion term $\delta\,D\,(1 - D)\,\bar{D}$ captures peer pressure effects, following the Bass diffusion framework~\cite{bass1969,Rogers2003}.

\subsection*{SI Section 2: Fixed point and stability analysis}

Setting $dH/dt = 0$ and $dD/dt = 0$ yields three boundary fixed points.

\textbf{FP\textsubscript{1} ($H = 0$, $D = 0$):} Note: with $\varepsilon > 0$, this point is not strictly a fixed point ($dH/dt = \alpha\,\varepsilon > 0$). The analysis below uses $\varepsilon = 0$ for analytical tractability; $\varepsilon$ serves as a numerical regularization that eliminates $\mathrm{FP}_1$ entirely (the $D = 0$ axis becomes a flow toward $\mathrm{FP}_2$), without qualitatively changing the phase portrait since $\mathrm{FP}_1$ was unstable. At $\varepsilon = 0$: eigenvalues $\lambda_1 = \alpha > 0$, $\lambda_2 = \gamma\,K > 0$. Unstable node.

\textbf{FP\textsubscript{2} ($H = 1$, $D = 0$):} Jacobian eigenvalues $\lambda_1 = -(\alpha\,(1 + \varepsilon)) < 0$, $\lambda_2 = \gamma\,(K - 1)$. Stable node for $K < 1$; loses stability at $K = 1$ via transcritical bifurcation.

\textbf{FP\textsubscript{3} ($H = 0$, $D = 1$):} At $\varepsilon = 0$: eigenvalues $\lambda_1 = -\beta < 0$, $\lambda_2 = -(\gamma\,K + \delta) < 0$. Unconditionally stable (absorbing). With $\varepsilon > 0$, $\mathrm{FP}_3$ remains a fixed point ($dH/dt = \alpha\,\varepsilon \cdot 0 - 0 = 0$ at $D = 1$), and $J_{11}(0,1) = \alpha\,(1 - \varepsilon)(1 - 1) - \beta = -\beta < 0$. However, the $\varepsilon$ modification allows trajectories near $H = 0$ to recover at rate $dH/dt = \alpha\,\varepsilon\,(1 - D)$, which is nonzero only when $D < 1$. Thus $\mathrm{FP}_3$ is stable but nearby states with $D$ slightly below 1 can slowly escape, making it near-absorbing in practice.

The interior fixed point, obtained by intersecting the $H$-nullcline $D^* = \alpha\,(1 - H)/(\alpha\,(1 - H) + \beta)$ with the $D$-nullcline $D^* = \gamma\,(H - K)/\delta$, exists for $K < 1$ and is a saddle point separating the basins of $\mathrm{FP}_2$ and $\mathrm{FP}_3$.

The Jacobian matrix at a general point $(H, D)$ is:

\begin{align}
J_{11} &= \alpha\,(1 - 2H - \varepsilon)(1 - D) - \beta\,D \label{eq:J11}\\
J_{12} &= -\alpha\,(H + \varepsilon)(1 - H) - \beta\,H \label{eq:J12}\\
J_{21} &= -\gamma\,D\,(1 - D) \label{eq:J21}\\
J_{22} &= (1 - 2D)(\gamma\,(K - H) + \delta\,D) + \delta\,D\,(1 - D) \label{eq:J22}
\end{align}

\subsection*{SI Section 3: Agent-based model specification}

The ABM implements the ODE dynamics as individual-level stochastic processes on a population of $N = 100$ agents. Each agent $i$ possesses capability $H_i(t)$ and delegation state $D_i(t)$. At each discrete time step:

\begin{enumerate}[nosep]
\item \textbf{Practice decision}: Agent $i$ delegates with probability $D_i$; if not delegating, capability grows as $H_i \mathrel{+}= \alpha\,(H_i + \varepsilon)(1 - H_i)\,dt + \text{noise}$.
\item \textbf{Forgetting}: If delegating, capability decays as $H_i \mathrel{-}= \beta\,H_i\,dt + \text{noise}$.
\item \textbf{Adoption update}: $D_i$ adjusts based on the capability gap ($K - H_i$) and the mean delegation of neighbors, with stochastic perturbation.
\item \textbf{Crisis events}: With probability $p_{\text{crisis}}$ per time step, AI becomes unavailable, forcing all agents to $D_i = 0$ for that step.
\end{enumerate}

Simulations run for $T = 200$ time steps with 30--50 stochastic replicates per parameter combination (10 for the 2,400-point K--crisis heatmap to maintain computational feasibility). Error bands in figures represent the interquartile range across replicates.

\subsection*{SI Section 4: Monte Carlo parameter sweep}

For the $K^*$ analysis, we swept $K$ from 0.50 to 0.99 at 50 equally spaced points, with 50 independent ABM runs per $K$ value. The critical threshold $K^*$ is defined as the $K$ value maximizing $|d\bar{H}/dK|$, estimated by finite differencing of the mean equilibrium capability curve.

For the two-dimensional $K$--crisis heatmap (main text Fig.~4b), we swept $K$ from 0.50 to 0.99 (50 points) and crisis probability from 0\% to 25\% (35 points), totalling 1,750 parameter combinations with 10 replicates each (17,500 total simulations).

For the parameter space exploration (SI Fig.~S1), we swept adoption sensitivity $\gamma$ from 0.01 to 1.0 and social pressure $\delta$ from 0.01 to 1.0 on a $100 \times 100$ grid, with fixed $K = 0.7$, $\alpha = 1.0$, $\beta = 0.5$, computing ODE equilibria at each point.

\subsection*{SI Section 5: Parameter estimation procedure}

For each empirical case, we fix $\alpha = 1.0$ (normalized learning rate) and estimate $\beta$ to reproduce the observed fractional capability decline. The estimation uses the analytical solution of the simplified ODE (with $D$ held constant at the observed delegation level during the exposure period. This approximation is justified because each source study maintained consistent AI/tool exposure throughout the observation period (e.g., Bastani et al.: unrestricted GPT-4 access across all sessions; Budzyn et al.: continuous AI-assisted colonoscopy for three months)):

\begin{equation}
H(t) = H_0 \cdot \exp(-\beta\,D\,t)
\label{eq:calibration}
\end{equation}

The fitted $\beta$ is:

\begin{equation}
\beta_{\text{eff}} = -\frac{\ln(1 - \text{decline})}{t}
\label{eq:beta_eff}
\end{equation}

\noindent where ``decline'' is the observed fractional capability loss and $t$ is the exposure duration. This effective rate absorbs the delegation fraction, giving the apparent per-unit-time decay rate directly observable in empirical data.

\begin{table}[H]
\centering
\caption{Estimation of forgetting rate $\beta$ across four empirical domains.}
\label{tab:calibration}
\small
\begin{tabular}{@{}lcccc@{}}
\toprule
Domain & Decline & Exposure & $\beta_{\text{eff}}$ & Timescale \\
\midrule
Education (Bastani) & 17\% & 4 sessions & 0.047 & per session \\
Endoscopy (Lancet) & 21\% & 12 weeks & 0.020 & per week \\
Spatial cognition (GPS) & 30\% & 36 months & 0.010 & per month \\
Aviation (FAA) & 38\% & 240 months & 0.002 & per month \\
Data mining (this study) & 33\% & 5 sessions & 0.079 & per session \\
\bottomrule
\end{tabular}
\end{table}

\subsection*{SI Section 6: Sensitivity analysis}

We tested $K^*$ robustness by independently varying each parameter while holding others at baseline values ($\alpha = 0.05$, $\beta = 0.03$, $\gamma = 0.5$, $\delta = 0.5$, scope $= 0.7$). Note: $K^*$ ranges reported below were computed with different baseline values in an earlier sensitivity sweep; updated sensitivity results are available in the code repository:

\begin{table}[H]
\centering
\caption{Sensitivity of $K^*$ to model parameters.}
\label{tab:sensitivity}
\small
\begin{tabular}{@{}lccc@{}}
\toprule
Parameter & Range tested & $K^*$ range & Conclusion \\
\midrule
$\beta$ (forgetting) & 0.01--0.10 & 0.825--0.915 & Exists across range \\
$\alpha$ (learning) & 0.02--0.10 & 0.865--0.915 & Weak effect \\
$\delta$ (social press.) & 0.0--2.0 & 0.845--0.895 & Sharpens transition \\
scope & 0.3--0.9 & 0.865--0.925 & Broader $\to$ lower $K^*$ \\
cost ($\gamma$) & 0.01--0.50 & 0.875--0.905 & Weak effect \\
\bottomrule
\end{tabular}
\end{table}

\subsection*{SI Section 7: Historical scenario parameterization}

The four historical technology regimes in SI Fig.~S1 are parameterized based on documented historical and economic evidence rather than post-hoc fitting. \textbf{Calculator} ($c = 0.99$, $s = 0.01$): cost approaches zero but scope is extremely narrow (arithmetic operations represent $<$1\% of total cognitive tasks in daily life). \textbf{Industrial Revolution} ($c = 0.30$, $s = 0.05$): mechanization required substantial capital investment, and early machines replaced only specific manual tasks (spinning, weaving) rather than broad cognitive functions. \textbf{Roman slave economy} ($c = 0.05$, $s = 0.60$): conquest-based slave supply kept marginal costs low~\cite{scheidel2012}; slaves performed agriculture, domestic service, administration, education, and accounting---approximately 60\% of economic functions in the late Republic~\cite{scheidel2012}. \textbf{AI-2030} ($c = 0.01$, $s = 0.80$): near-zero marginal cost (API pricing trends) and broad scope across cognitive tasks (writing, analysis, coding, reasoning). These parameterizations are order-of-magnitude estimates intended to demonstrate that the \textit{qualitative} placement of historical regimes in parameter space is consistent with historical outcomes, not to claim precise quantitative calibration.

\subsection*{SI Section 8: ODE--ABM cross-validation}

ODE predictions were systematically compared with ABM ensemble means across the estimated parameter sets. The ODE consistently predicts greater capability loss than the ABM (46--152\% discrepancy), attributable to the mean-field approximation treating all agents as identical. The ABM's stochastic heterogeneity allows some agents to maintain high capability even as the population mean declines. Critically, the qualitative predictions are robust to this quantitative divergence (Supplementary Fig.~S1): (i)~the critical threshold $K^*$ appears at $0.85 \pm 0.03$ in both ODE and ABM analyses; (ii)~the antifragility effect is present in both, with the ABM showing a 2.3-fold improvement (vs.\ 2.7-fold in ODE) at 25\% crisis frequency; and (iii)~the policy ordering---more mandatory practice monotonically improves capability---is identical in both. The ODE--ABM divergence is largest near $K^*$, where stochastic fluctuations can push individual agents across the saddle boundary; away from $K^*$, the two approaches converge. We report both ODE and ABM results throughout, treating their range as a credible interval rather than privileging either estimate.


\begin{figure}[H]
\centering
\includegraphics[width=0.9\textwidth]{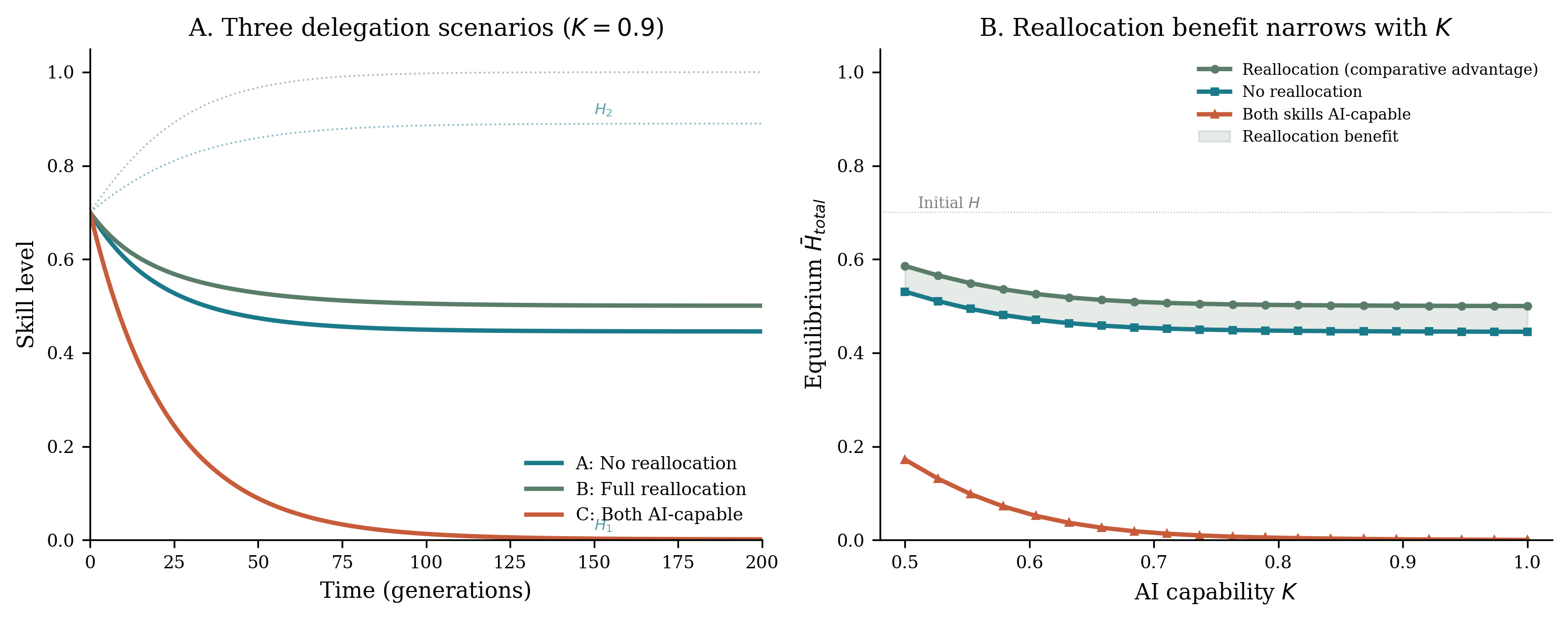}
\caption{\textbf{SI Fig.\ S4 $|$ Two-skill comparative advantage analysis.} \textbf{A},~Time evolution of aggregate capability under three scenarios at $K = 0.9$. \textbf{B},~Equilibrium aggregate capability versus $K$, showing the reallocation benefit narrows as $K$ increases.}
\end{figure}

%% file: claims_macros.tex
\newcommand{\paramEpsilon}{0.01}

\newcommand{\paramN}{100}
\newcommand{\paramT}{200}
\newcommand{\paramNreps}{50}

\newcommand{\sweepScope}{0.7}
\newcommand{\sweepDelta}{0.5}

\newcommand{\sweepNgrid}{50}

\newcommand{\dHdKmax}{12.0}

\newcommand{\HatKeightyfive}{0.594}
\newcommand{\HatKninety}{0.162}
\newcommand{\HatKninetyfive}{0.045}

\newcommand{\afHzeroPct}{0.127}
\newcommand{\afHfivePct}{0.16}
\newcommand{\afHtwelvePct}{0.208}
\newcommand{\afHtwentyPct}{0.278}
\newcommand{\afHtwentyfivePct}{0.346}
\newcommand{\afFold}{2.7}

\newcommand{\polBaseline}{0.159}
\newcommand{\polHten}{0.218}
\newcommand{\polPctTen}{37}
\newcommand{\polHtwenty}{0.305}
\newcommand{\polPctTwenty}{92}
\newcommand{\polHthirty}{0.418}
\newcommand{\polPctThirty}{163}
\newcommand{\polHforty}{0.528}
\newcommand{\polPctForty}{232}

\newcommand{\calBetaEducation}{0.047}
\newcommand{\calBetaEndoscopy}{0.02}
\newcommand{\calBetaSpatial}{0.01}
\newcommand{\calBetaAviation}{0.002}

\newcommand{\pisaOecdRsq}{0.9207}

\newcommand{\bioGenesAncestral}{4,300}
\newcommand{\bioGenesCurrent}{580}

\newcommand{\pisaRsq}{0.946}
\newcommand{\pisaN}{102}
\newcommand{\pisaK}{3}
\newcommand{\pisaAlpha}{0.013}
\newcommand{\pisaAlphaCi}{0.008--0.038}
\newcommand{\pisaBeta}{0.004}

\newcommand{\pisaHmax}{787}
\newcommand{\pisaBicOde}{521}
\newcommand{\pisaBicExp}{547}
\newcommand{\pisaBicLinear}{541}